\documentclass[aps,prl,twocolumn,showpacs,amsmath,amssymb, nofootinbib]{revtex4-1}
\usepackage{epsfig}
\usepackage{graphicx}
\usepackage{dcolumn}
\usepackage{bm}
\usepackage{ltablex,booktabs}
\usepackage{overpic}
\usepackage{subfigure}
\usepackage{siunitx}
\usepackage{float}
\usepackage{ulem}
\usepackage{relsize}
\usepackage{color}
\usepackage{amsmath}
\usepackage{mathcomp}
\usepackage{mathrsfs}
\usepackage{multirow}
\usepackage{rotating}
\usepackage{amssymb}
\usepackage{gensymb}
\usepackage{amsmath}
\usepackage{tabularx}
\usepackage{lineno}

\usepackage[bookmarksnumbered, pdfstartview=FitH,colorlinks,urlcolor=blue, citecolor=blue,linkcolor=blue] {hyperref}

\begin{document}
\normalsize
\parskip=5pt plus 1pt minus 1pt

\title{\boldmath Study of  $D^{+} \to K_{S}^{0}K^{*}(892)^{+}$ in $D^{+} \to K_{S}^{0} K_{S}^{0} \pi^{+}$  }

\author{
\begin{small}
\begin{center}
M.~Ablikim$^{1}$, M.~N.~Achasov$^{4,c}$, P.~Adlarson$^{76}$, O.~Afedulidis$^{3}$, X.~C.~Ai$^{81}$, R.~Aliberti$^{35}$, A.~Amoroso$^{75A,75C}$, Q.~An$^{72,58,a}$, Y.~Bai$^{57}$, O.~Bakina$^{36}$
% ...
\\
\vspace{0.2cm}
(BESIII Collaboration)\\
\vspace{0.2cm} {\it
$^{1}$ Institute of High Energy Physics, Beijing 100049, People's Republic of China\\
% ...
$^{3}$ Bochum  Ruhr-University, D-44780 Bochum, Germany\\
$^{4}$ Budker Institute of Nuclear Physics SB RAS (BINP), Novosibirsk 630090, Russia\\
% ...
$^{35}$ Johannes Gutenberg University of Mainz, Johann-Joachim-Becher-Weg 45, D-55099 Mainz, Germany\\
$^{36}$ Joint Institute for Nuclear Research, 141980 Dubna, Moscow region, Russia\\
% ...
$^{57}$ Southeast University, Nanjing 211100, People's Republic of China\\
$^{58}$ State Key Laboratory of Particle Detection and Electronics, Beijing 100049, Hefei 230026, People's Republic of China\\
% ...
$^{72}$ University of Science and Technology of China, Hefei 230026, People's Republic of China\\
% ...
$^{75}$ University of Turin and INFN, (A)University of Turin, I-10125, Turin, Italy; (B)University of Eastern Piedmont, I-15121, Alessandria, Italy; (C)INFN, I-10125, Turin, Italy\\
$^{76}$ Uppsala University, Box 516, SE-75120 Uppsala, Sweden\\
% ...
$^{81}$ Zhengzhou University, Zhengzhou 450001, People's Republic of China\\
\vspace{0.2cm}
$^{a}$ Deceased\\
% ...
$^{c}$ Also at the Novosibirsk State University, Novosibirsk, 630090, Russia\\
% ...
}
\end{center}
\vspace{0.4cm}
\end{small}
}
\noaffiliation{}

\date{\today}
\begin{abstract}
  Using a data sample of $e^+e^-$ collisions corresponding to an integrated luminosity of 7.93 $\rm fb^{-1}$ collected with the BESIII detector at the center-of-mass energy 3.773~GeV, we perform the first amplitude analysis of the decay $D^{+} \to K_{S}^{0} K_{S}^{0} \pi^{+}$. The absolute branching fraction of  $D^{+} \to K_{S}^{0}K_{S}^{0} \texorpdfstring{\pi}{pi}^{+}$ is measured to be $(2.97 \pm 0.09_{\rm stat.} \pm 0.05_{\rm syst.})\times10^{-3}$. The dominant intermediate  process is $D^{+} \to K_{S}^{0}K^{*}(892)^{+}$, whose branching fraction is determined to be $(8.72 \pm 0.28_{\rm stat.} \pm 0.15_{\rm syst.}) \times 10^{-3}$, including all the $K^*(892)^+$ decays.
\end{abstract}
\maketitle
\section{I.~Introduction}
The masses of the charmed mesons, approximately 2~GeV/c$^{2}$, fall within a
region where non-perturbative Quantum Chromodynamics~(QCD) is no longer negligible.
This presents a challenge for studying the hadronic transitions of charmed meson.
The lightest charmed mesons $D^{0(+)}$ can only decay through weak interactions, 
and the related amplitudes are dominated by
two-body processes, i.e.~$D \to VP$, $D \to PP$, $D \to SP$ and $D \to VV$ decays,
where $V$, $S$, and $P$ denote vector, scalar and pseudoscalar mesons, respectively.
In particular, the
$D \to VP$ decays offer cleaner opportunities than other two-body processes
for clarifying the nonperturbative mechanism of charmed meson decays~\cite{3, 4, Cheng:VP2010}.
Compared to the $D \to SP$ decay,
the $VP$ system presents a more well-defined quark content of the vector meson than that of the scalar state, whose
 quark content is still a matter of controversy and is further complicated by significant
long-distance rescattering effects~\cite{BCKa03, BCKa0, BCKa02}.
The $VP$ system also can be more favorably described by the theoretical framework than
the $PP$ and $VV$ systems, due to the multiple-amplitude composition of some $P$ states~\cite{4, Cheng:VP2010} and
the polarization in $VV$ systems~\cite{zhaoqDtoVV}.

\begin{figure*}[htbp]
  \centering
  \includegraphics[width=6.6cm]{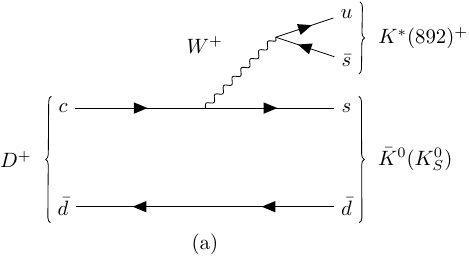}
  \includegraphics[width=6.6cm]{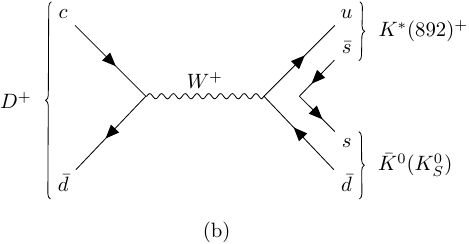}\\
  \includegraphics[width=6.6cm]{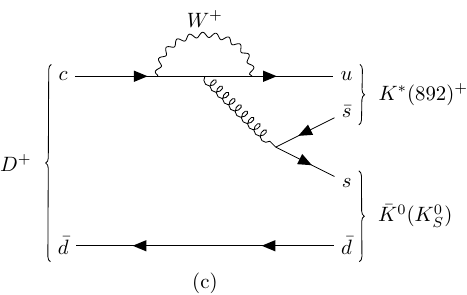}
  \includegraphics[width=6.6cm]{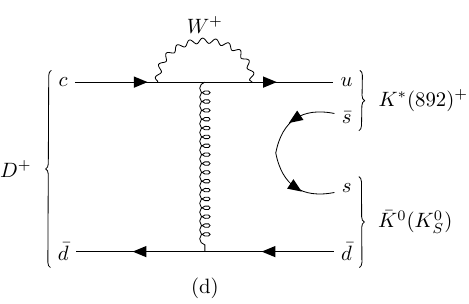}\\
  \caption{Topological diagrams contributing to the decay $D^+ \to K_S^0K^*(892)^+$ with (a) color allowed external $W$-emission tree diagram, (b) $W$-annihilation diagram, (c) color suppressed external $W$-emission QCD-penguin diagram and (d) QCD-penguin exchange diagram. }
  \label{FMT}
\end{figure*}

While most of the theoretical predicted branching fractions~(BFs) of $D \to VP$ decays are consistent
with experimental measurements, that for $D^+ \to K_S^0 K^*(892)^+$ shows inconsistency. 
The $D^+ \to K_S^0 K^*(892)^+$ decay is one of the most important
singly-Cabbibo-suppressed~(SCS) $D \to VP$ decays~\cite{kskpi0,xxh}, which can be
mediated via color-favored, $W$-annihilation, and penguin diagrams~\cite{4} as shown in Fig.~\ref{FMT}.
The  predicted and measured values of the BF $\displaystyle\mathcal{B}(D^+ \to K_S^0 K^*(892)^+ )$
are listed in Table~\ref{Consistentency_Check}, where predicted BFs of the decay
$D^+ \to K_S^0K^*(892)^+ $ are from the pole model~\cite{1}, the factorization-assisted
topological-amplitude (FAT [mix]) approach with $\rho-\omega$ mixing~\cite{2}, the
topological diagram approach (TDA) with only tree level amplitude (denoted as TDA [tree])~\cite{3}, including QCD-penguin amplitudes (denoted as TDA [QCD-penguin])~\cite{4}  and the updated analysis of the two-body $D \to VP$ decays within the framework of the TDA~\cite{06316}.

An amplitude analysis of decays with multi-body final states is performed to extract the information of subprocess via intermediate resonance. 
The BESIII collaboration has measured the BF of  $D^+ \to K_S^0 K^*(892)^{+}$ to
be $(8.69 \pm 0.40_{\rm stat.} \pm 0.64_{\rm syst.} \pm 0.51_{\rm Br.})\times10^{-3}$
in the amplitude analysis of $D^+ \to K^+ K_S^0 \pi^{0}$~\cite{xxh} using a data sample corresponding to an integrated luminosity of 2.93
$\rm fb^{-1}$ at the center-of-mass energy 3.773~GeV, where the third
uncertainty is due to the quoted uncertainty of
$\mathcal{B}(D^{+} \to K^{+}K_{S}^{0}\pi^{0})$~\cite{kskpi0}. This result differs
from the theoretical predictions in Refs.~\cite{2,3,4} by about 4$\sigma$.
Therefore, a more precise measurement of $D^+\to K_S^0 K^*(892)^+$ is needed to
provide a more stringent test of the models and to deepen our understanding
of the dynamics of charmed meson decays.
A clearer environment is expected to be offered in the $D^{+} \to K_{S}^{0}K_{S}^{0}\pi^{+}$ decay 
without the interference from $D^+\to K^{+} K^*(892)^0$ in $D^{+} \to K^{+}K_{S}^{0}\pi^{0}$.

The $D^{+} \to K_{S}^{0}K_{S}^{0}\pi^{+}$ decay
has been previously observed  by the BESIII collaboration with a BF of
$\displaystyle\mathcal{B}(D^+ \to K_S^0 K_S^0 \pi^{+} )= (2.70 \pm 0.05_{\rm stat.} \pm 0.12_{\rm syst.}) \times 10^{-3}$~\cite{wangyue}. 
 Using a data sample of $e^+e^-$ collisions corresponding to an integrated luminosity of 7.93 $\rm fb^{-1}$~\cite{zhanghan} collected by the BESIII detector~\cite{Ablikim:2009aa} at the center-of-mass energy 3.773~GeV, we performed the first amplitude analysis and the absolute BF measurement of the SCS decay $D^{+} \to K^{0}_{S}K_{S}^{0}\pi^{+}$.  Charge-conjugate states are implied throughout this paper, allowing for the signal-side decay $D^- \to K_S^0 K_S^0 \pi^-$.

\begin{table}[hbtp]
\begin{center}
  \caption{Theoretical predictions of $\displaystyle\mathcal{B}(D^+ \to K_S^0 K^*(892)^+ )$ and the previous experimental measurement.}

	\begin{tabular}{l c}
	\hline
	\hline
	Model     &$\displaystyle\mathcal{B}(D^+ \to K_S^0K^*(892)^+)$ $(\times 10^{-3})$     \\
	\hline
	Pole~\cite{1}        	&$6.2 \pm 1.2$  \\
	FAT [mix]~\cite{2}			&5.5\\
	TDA [tree]~\cite{3} 			&$5.02 \pm 1.31$ \\
	TDA [QCD-penguin]~\cite{4} 			&$4.90 \pm 0.21$ \\
	TDA ~\cite{06316}        &$7.90 \pm 0.25$\\
    \hline
    BESIII measurement~\cite{xxh} &$8.69 \pm 0.40_{\rm stat.} \pm 0.64_{\rm syst.} \pm 0.51_{\rm BF.}$\\
       
	\hline
	\hline
	\end{tabular}
	\label{Consistentency_Check}
\end{center}
\end{table}

\section{II.~Detector and Monte Carlo simulation}
The BESIII detector~\cite{Ablikim:2009aa} records symmetric $e^+e^-$ collisions provided by the BEPCII  storage ring~\cite{Yu:IPAC2016-TUYA01} in the center-of-mass energy range from 2.00 to 4.95~GeV, with a peak luminosity of $1 \times 10^{33}\;\text{cm}^{-2}\text{s}^{-1}$  achieved at $\sqrt{s} = 3.77\;\text{GeV}$. 
BESIII has collected large data samples in this energy region~\cite{Ablikim:2019hff,EcmsMea,EventFilter}. The cylindrical core of the BESIII detector covers 93\% of the full solid angle and consists of a helium-based multilayer drift chamber~(MDC), a plastic scintillator time-of-flight system~(TOF), and a CsI(Tl) electromagnetic calorimeter~(EMC), which are all enclosed in a superconducting solenoidal magnet
providing a 1.0~T magnetic field. The solenoid is supported by an octagonal flux-return yoke with resistive plate counter muon identification modules interleaved with steel. The charged-particle momentum resolution at $1~{\rm GeV}/c$ is $0.5\%$, and the ${\rm d}E/{\rm d}x$ resolution is $6\%$ for electrons from Bhabha scattering. The EMC measures photon energies with a resolution of $2.5\%$ ($5\%$) at $1$~GeV in the barrel (end cap) region. The time resolution in the TOF barrel region is 68~ps, while that in the end cap region was 110~ps.  The end cap TOF system was upgraded in 2015 using multigap resistive plate chamber technology, providing a time resolution of 60~ps, which benefits 63\% of the data used in this analysis~\cite{etof}.

Simulated inclusive Monte Carlo (MC) samples are produced with a {\sc geant4}-based~\cite{geant4} MC simulation package, which includes the geometric description of the BESIII detector~\cite{NST33142} and the detector response, and are used to determine detection efficiencies and to estimate backgrounds. The simulation models the beam energy spread and initial state radiation~(ISR) in the $e^+e^-$ annihilations with the generator {\sc kkmc}~\cite{ref:kkmc}. The inclusive MC samples consist of the production of $D\bar{D}$ pairs, the non-$D\bar{D}$ decays of the $\psi(3770)$, the ISR production of the $J/\psi$ and $\psi(3686)$ states, and the continuum processes incorporated in {\sc kkmc}. All particle decays are modelled with {\sc evtgen}~\cite{ref:evtgen} using BFs either taken from the Particle Data Group (PDG)~\cite{PDG}, when available, or otherwise estimated with {\sc lundcharm}~\cite{ref:lundcharm}. Final state radiation from charged final state particles is incorporated using {\sc photos}~\cite{photos}. In this work, two sets of MC samples are used. The phase space (PHSP) MC sample is generated with a uniform distribution in phase space for the decay $D^+ \to  K_S^0 K_S^0 \pi^+$, which is used to calculate the normalization factor of the probability density function (PDF) in the amplitude analysis. The signal MC sample, which is used to estimate the detection efficiencies, is generated  according to the amplitude analysis results.

\section{III.~Event selection}
The process $e^{+}e^{-} \to \psi(3770) \to D^{+}D^{-}$ allows studies of $D$ decays with a tag technique~\cite{MarkIII-tag, Ke:2023qzc}. There are two types of samples used with the tag technique: single tag~(ST) and
double tag~(DT). In the ST sample, only one $D^{-}$ meson is reconstructed through one of six hadronic decay modes: $D^-\to K^{+}\pi^{-}\pi^{-}$, $K^{+}\pi^{-}\pi^{-}\pi^{0}$, $K^{0}_{S}\pi^{-}$, $K_{S}^{0}\pi^{-}\pi^{0}$, $K_{S}^{0}\pi^{-}\pi^{-}\pi^{+}$, and $K^{+}K^{-}\pi^{-}$. In the DT sample, the signal $D^{+}$ is reconstructed through $D^{+} \to K^0_{S}K^0_{S}\pi^{+}$.

All charged tracks detected in the MDC must satisfy $|$cos$\theta|<0.93$, where $\theta$ is defined as the polar angle with respect to the $z$-axis, which is the symmetry axis of the MDC. For charged tracks not originating from $K_S^0$ decays, the distance of closest approach to the interaction point (IP) is required to be less than 1\,cm in the transverse plane ($|V_{xy}|$), and less than 10\,cm along the $z$-axis ($|V_{z}|$). Particle identification (PID) for charged tracks combines the dE/dx measurement in the MDC with the time of flight measurement of the TOF detector to define the likelihood function $\mathcal{L}(h)~(h=K,\pi)$ for each hadron ($h$) hypothesis. The charged kaons and pions are identified by comparing the likelihoods for the kaon and pion hypotheses, $\mathcal{L}(K)>\mathcal{L}(\pi)$ and $\mathcal{L}(\pi)>\mathcal{L}(K)$, respectively. 

Each $K_{S}^0$ candidate is reconstructed from two oppositely charged tracks satisfying \mbox{$|V_{z}|<$ 20~cm}. The two charged tracks are assigned as $\pi^+\pi^-$ without imposing PID. They are constrained to originate from a common vertex and are required to have an invariant mass $M_{\pi^+\pi^-}$ such that $|M_{\pi^{+}\pi^{-}} - m_{K_{S}^{0}}|<$ 12~MeV$/c^{2}$, where $m_{K_{S}^{0}}$ is the known $K^0$ mass~\cite{PDG}. The decay length of the $K^0_S$ candidate is required to be greater than twice its resolution.
 
Photon candidates are selected using  EMC showers. The deposited energy of each shower in the barrel region~($|\!\cos \theta|< 0.80$) and in the end-cap region~($0.86 <|\!\cos \theta|< 0.92$) must be greater than 25 MeV and 50 MeV, respectively. To exclude showers that originate from charged tracks, the angle subtended by the EMC shower and the position of the closest charged track at the EMC must be greater than 10 degrees as measured from the IP. The difference between the EMC time and the event start time is required to be within \mbox{[0, 700]~ns} to suppress electronic noise and showers unrelated to the event.

The $\pi^0$ candidates are reconstructed from photon pairs with invariant masses in the range $[0.115, 0.150]$~GeV/$c^{2}$, which corresponds to about three times the invariant mass resolution. We require that at least one photon comes from the barrel region of the EMC to improve the resolution. Furthermore, the $\pi^0$ candidates are constrained to the known $\pi^0$ mass~\cite{PDG} via a kinematic fit to improve their energy and momentum resolution.

Two variables, the beam-constrained mass $M_{\rm BC}$ and the energy difference $\Delta E$, are used to identify the $D^+$ mesons:
\begin{eqnarray}
\begin{aligned}
    M_{\rm BC}&=\sqrt{E^2_{\rm beam}/c^4-|\vec{p}_{D^+}|^2/c^2},\\
	\Delta{E} &= E_{D^+}-E_{\rm beam}, \label{eq:mbc}
\end{aligned}
\end{eqnarray}
where $E_{\rm beam}$ is the  beam energy, and $\vec{p}_{D^+}$ and $E_{D^+}$ are the total reconstructed momentum and energy of the $D^+$ candidate in the $e^+e^-$ rest frame, respectively. The $D^+$ signals  appear as a peak at the known $D^+$ mass~\cite{PDG} in the $M_{\rm BC}$ distribution and as a peak at zero in the $\Delta{E}$ distribution. The  selection criteria of $K_{S}^{0}$ and $\pi^{+}$ on the signal side are the same as those on the tag side. If multiple DT candidates exist in one event, the candidate with the minimum $\lvert \Delta E_{\rm sig} \rvert$ from the $D^+$ mesons is retained.  
\begin{table}[hbtp]
    \centering
    \caption{Requirements of $\Delta E_{\rm tag}$ for different $D^{-}$ tag modes. }
    \begin{tabular}{llllll ccc}
    \hline
    \hline
      &  $D^{-}$ decay  &$\Delta E_{\rm tag}$~(GeV)  \\
      \hline
      &  $ K^{+}\pi^{-}\pi^{-}$    &  $(-0.025, 0.024)$ \\
      &  $ K^{+}\pi^{-}\pi^{-}\pi^{0}$ &  $(-0.057, 0.046)$\\
      &  $ K_{S}^{0}\pi^{-}$ & $(-0.025, 0.026)$ \\
      &  $ K_{S}^{0}\pi^{-}\pi^{0}$& $(-0.062, 0.049)$ \\
      &  $ K_{S}^{0}\pi^{-}\pi^{-}\pi^{+}$ & $(-0.028, 0.027)$ \\
      & $ K^{+}K^{-}\pi^{-}$ & $(-0.024, 0.023)$ \\
  \hline
  \hline
    \end{tabular}
    \label{tagdeltaE}
\end{table}

\section{IV.~Background study}
\begin{figure}[htbp]
    \centering
    \includegraphics[width=7cm]{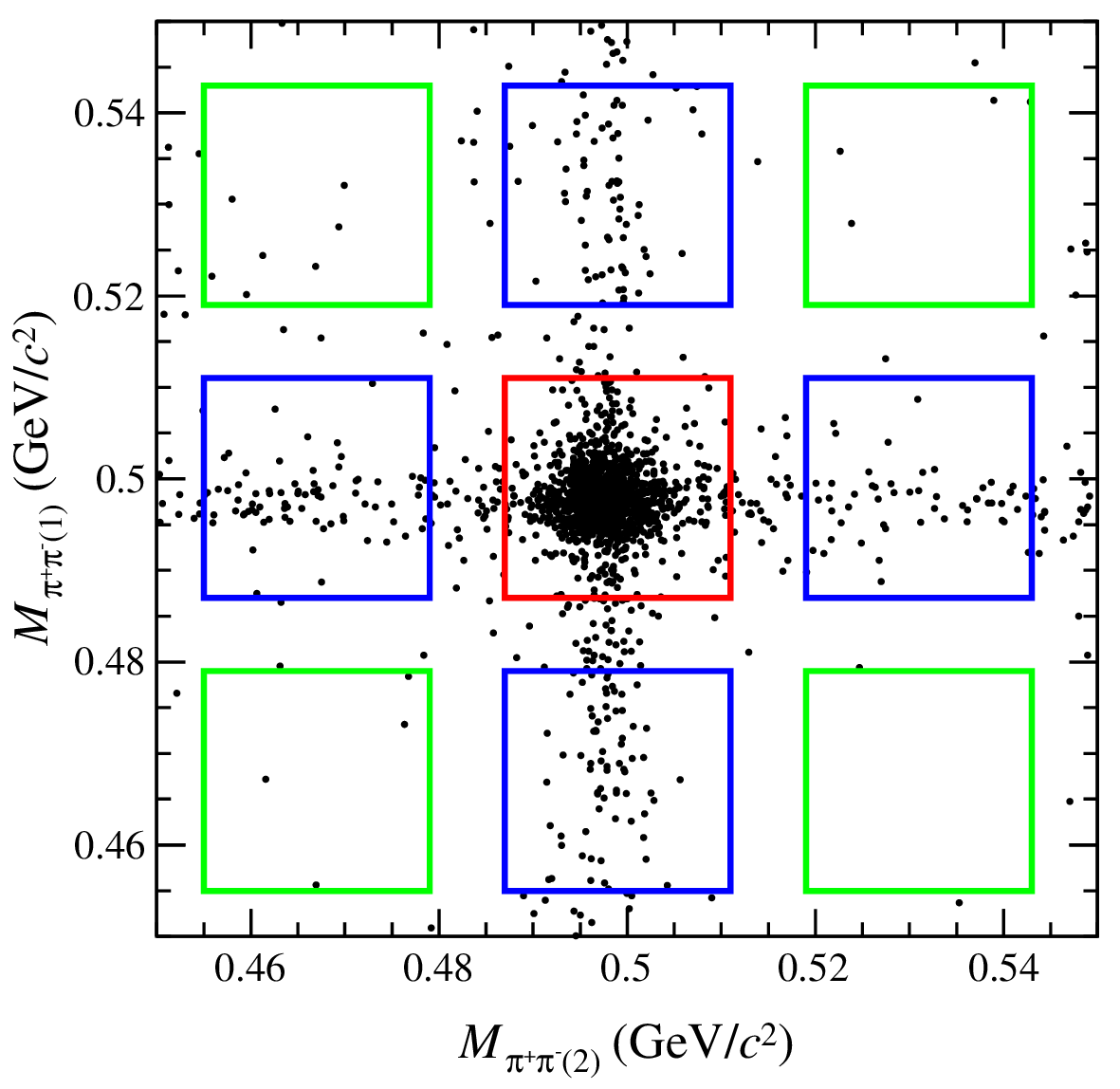}
    \caption{Distribution of $M_{\pi^+\pi^-(1)}$ versus $M_{\pi^+\pi^-(2)}$ for the $D^+ \to K_S^0 K_S^0 \pi^+$ candidate events in data. The red solid box denotes the 2D signal region. The blue (green) solid boxes indicate the 2D sideband 1 (2) regions.}
    \label{fig:2Dks}
\end{figure}

To increase the signal purity for amplitude analysis, we require $1.863 < M_{\rm BC} < 1.877$~GeV/$c^{2}$ for all  tag modes, and $1.865 < M_{\rm BC} <1.875$~GeV/$c^{2}$ and $-0.020 < \Delta E < 0.020$~GeV for the signal side. The boundaries of $\Delta E_{\rm tag}$ requirements are listed in Table~\ref{tagdeltaE}. Then,
we study the sources of background for $ D^+ \to K_S^0 K_S^0 \pi^+$ by analyzing the inclusive MC samples. The main background is the decay $D^+ \to K_S^0 \pi^+ \pi^- \pi^+$, since the combinatorial $\pi^+\pi^-$ pairs may accidentally satisfy the $K_S^0$ selection criteria and contribute as a peak around the $D^{+}$ mass in the $M_{\rm BC}$ distribution. This peaking background is estimated with events in the $K_S^0$ sideband region, defined as $0.020\!<\!|M_{\pi^+\pi^-}\!-\!m_{K_S^0}|\!<\!0.044$~GeV$/c^2$. The two dimensional (2D) $K_S^0$ signal region is defined as the square region with both $\pi^+\pi^-$ combinations lying in the 1D $K_S^0$ signal region. The 2D $K_S^0$ sideband 1 (2) regions are defined as the square regions with 1 (2) $\pi^+\pi^-$ combination(s) located in the 1D $K_S^0$ sideband regions and the rest in the 1D $K_S^0$ signal region.
Figure~\ref{fig:2Dks} shows the 2D distribution of $M_{\pi^+\pi^-(1)}$ versus $M_{\pi^+\pi^-(2)}$ for the $D^+ \to K_S^0 K_S^0 \pi^+$ candidates in data, where the two $K_S^0$ candidates are randomized.

\begin{figure*}[htbp]
  \centering

  \includegraphics[width=5cm]{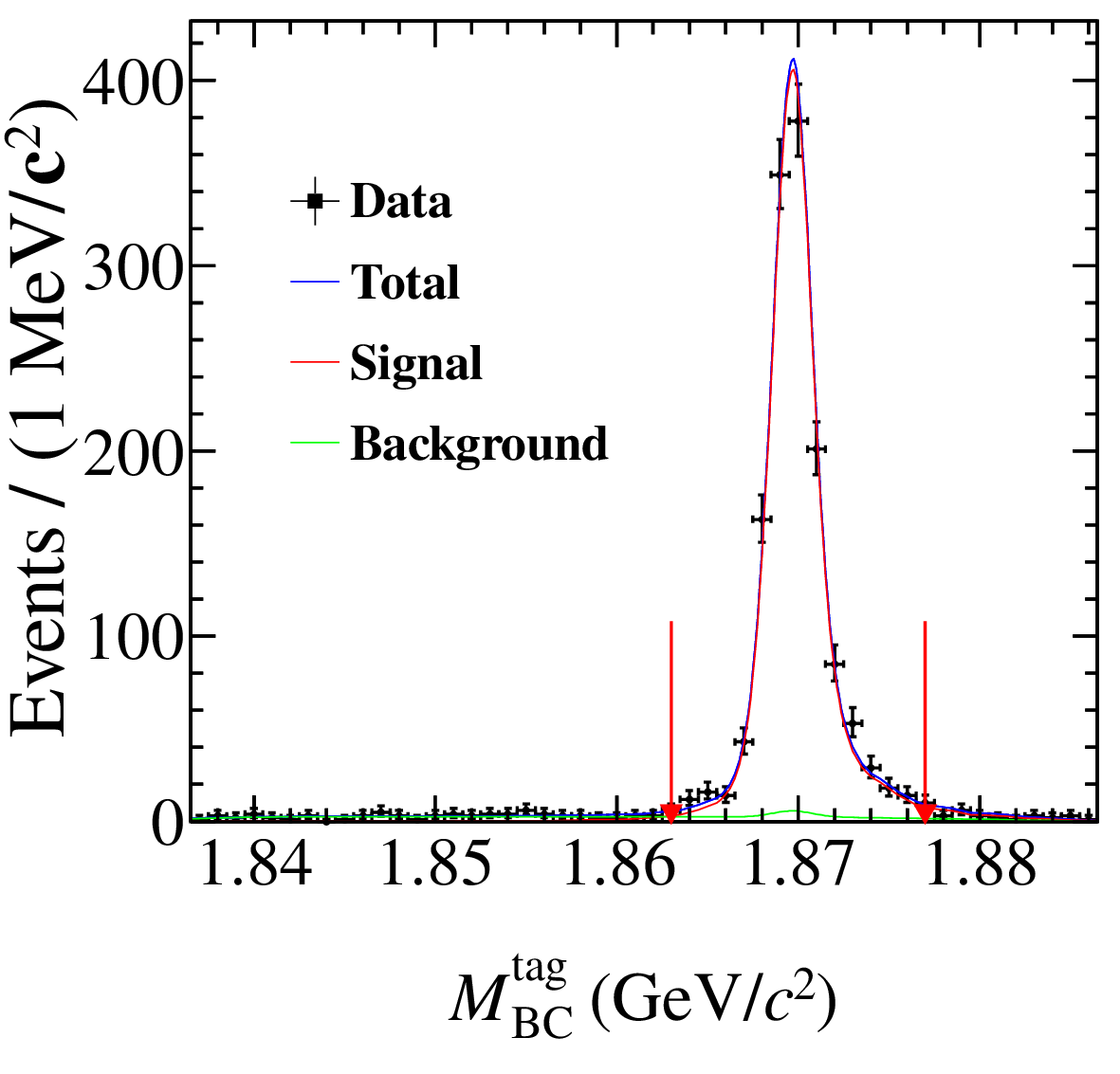} 
  \includegraphics[width=5cm]{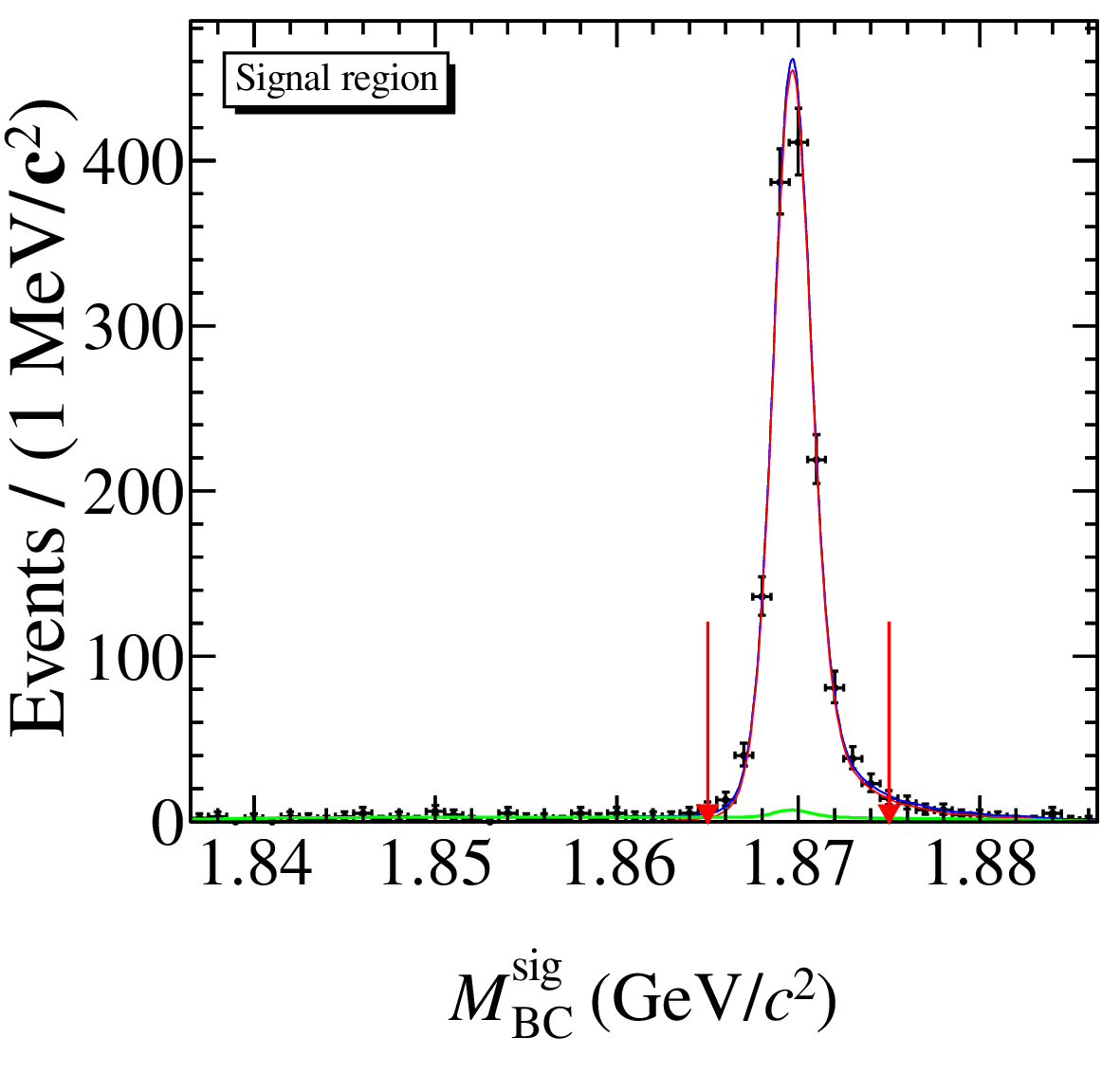} 
  \includegraphics[width=5cm]{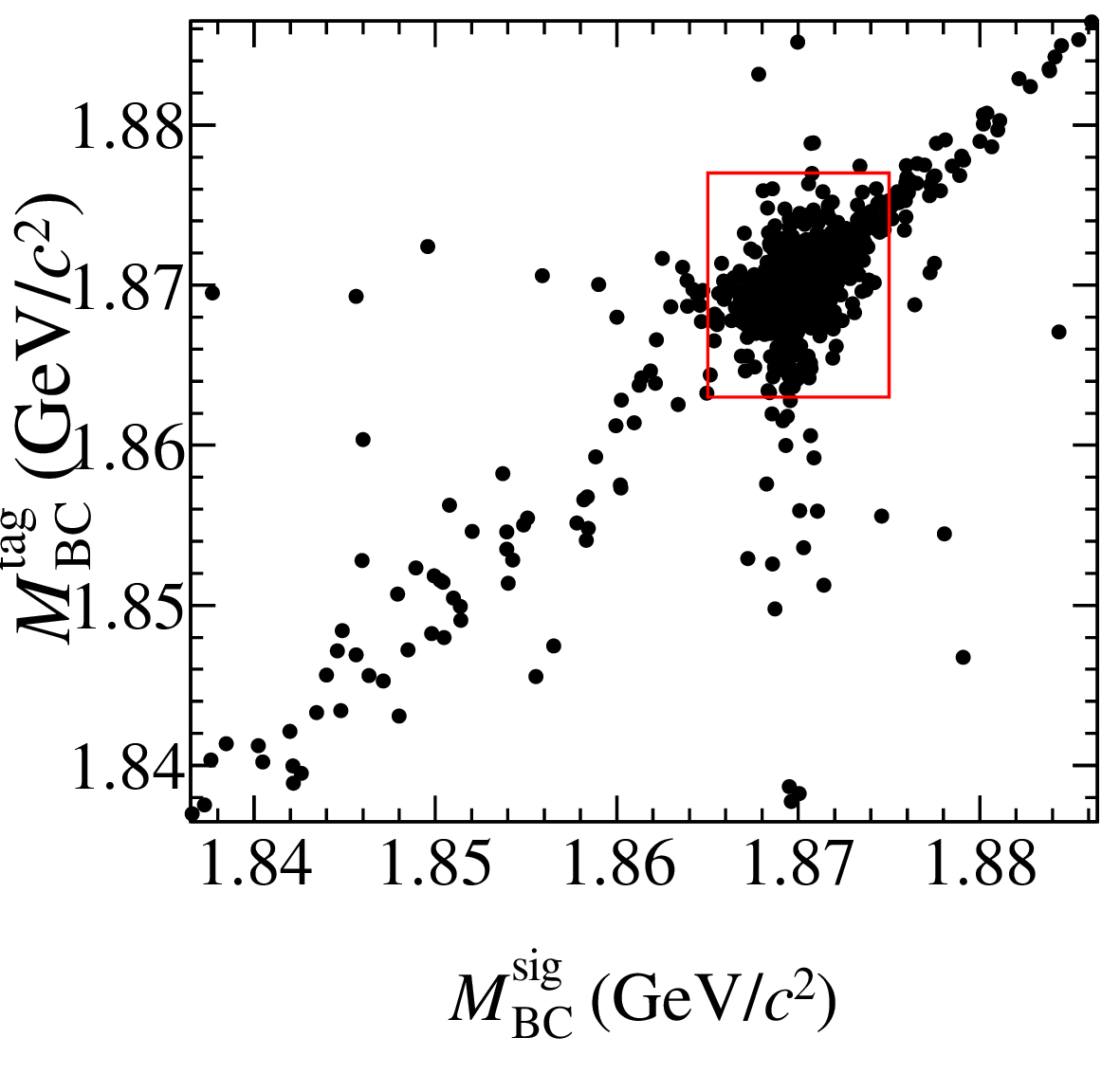}\\
  \includegraphics[width=5cm]{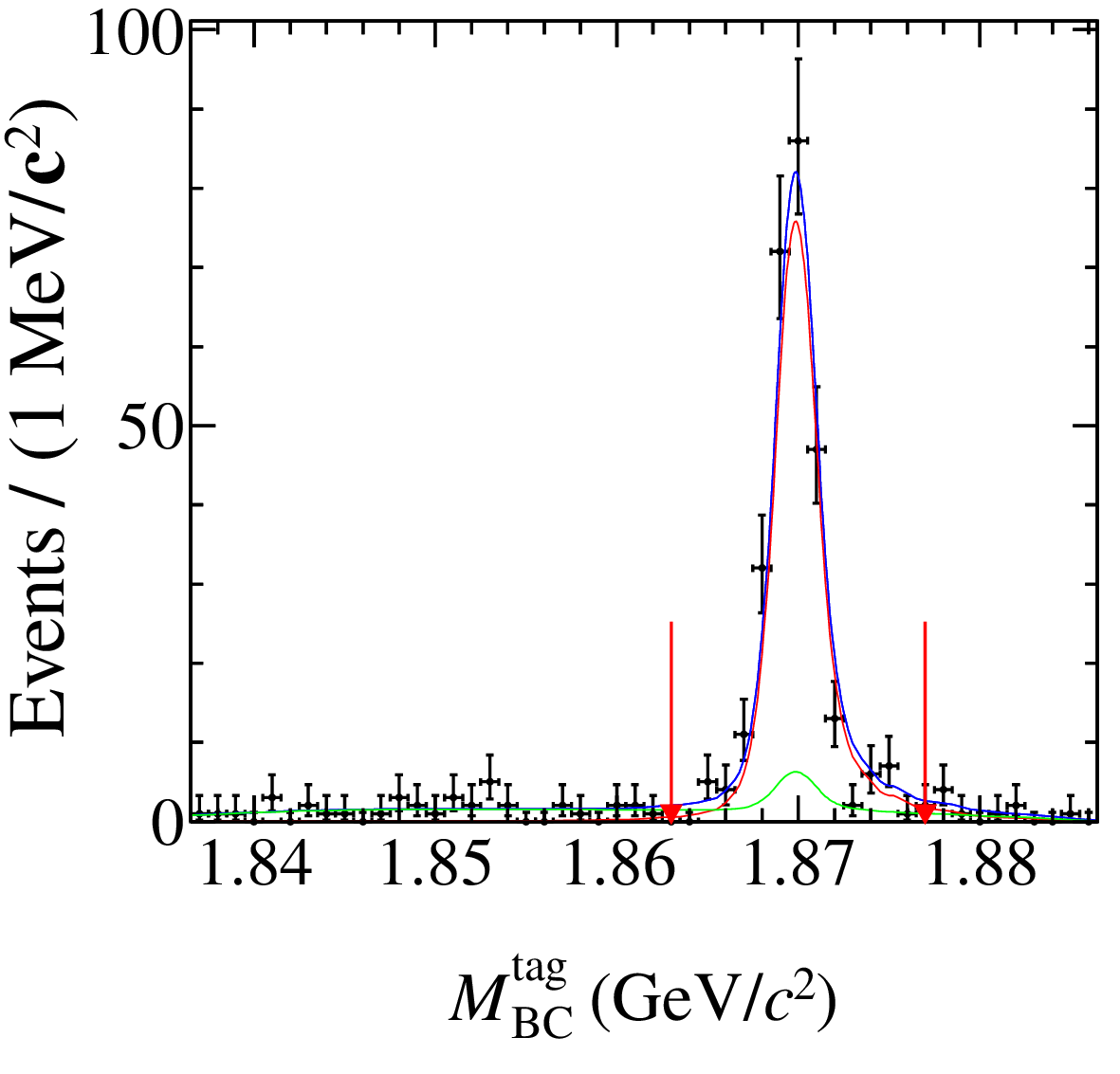}
  \includegraphics[width=5cm]{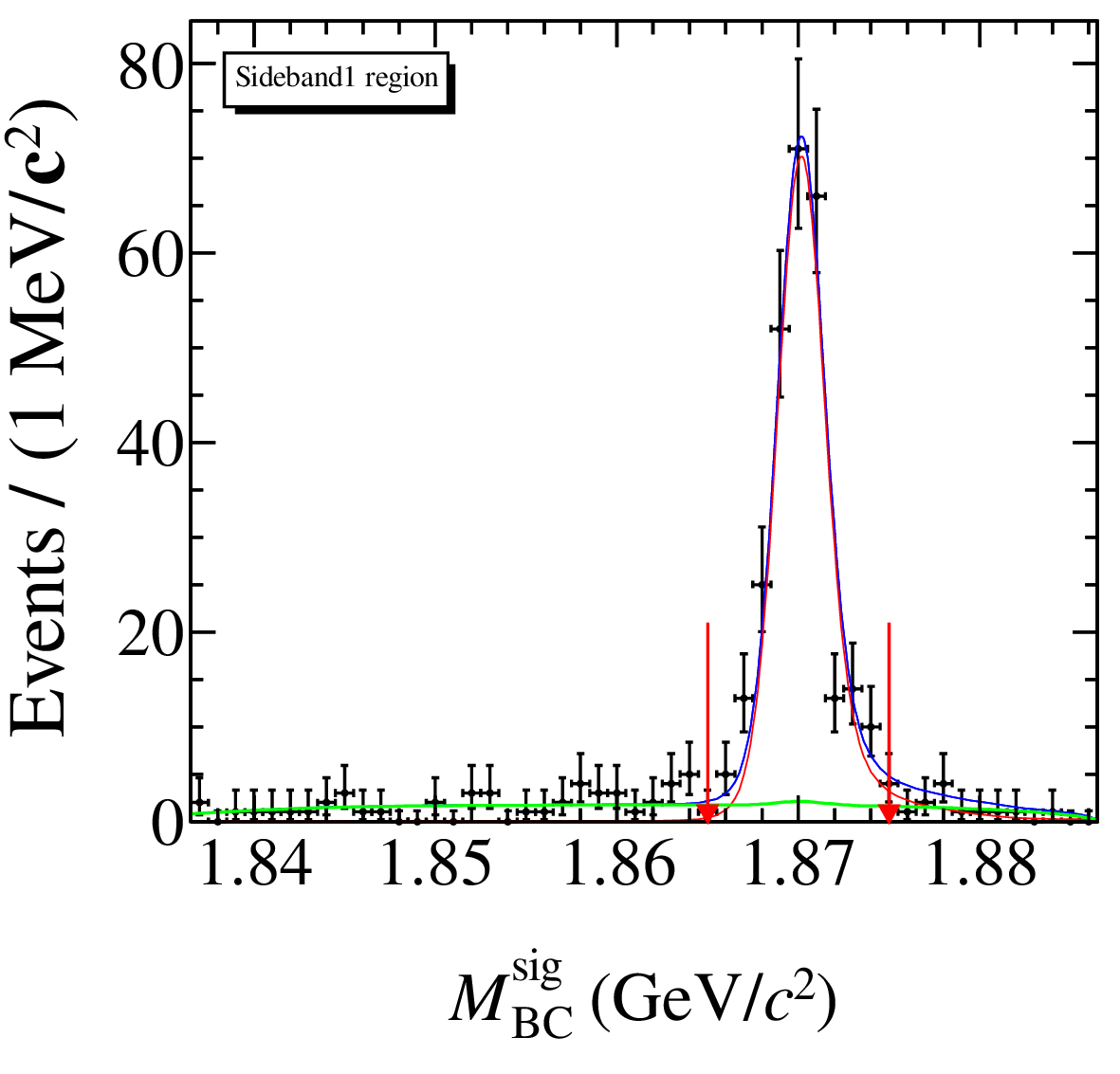}
  \includegraphics[width=5cm]{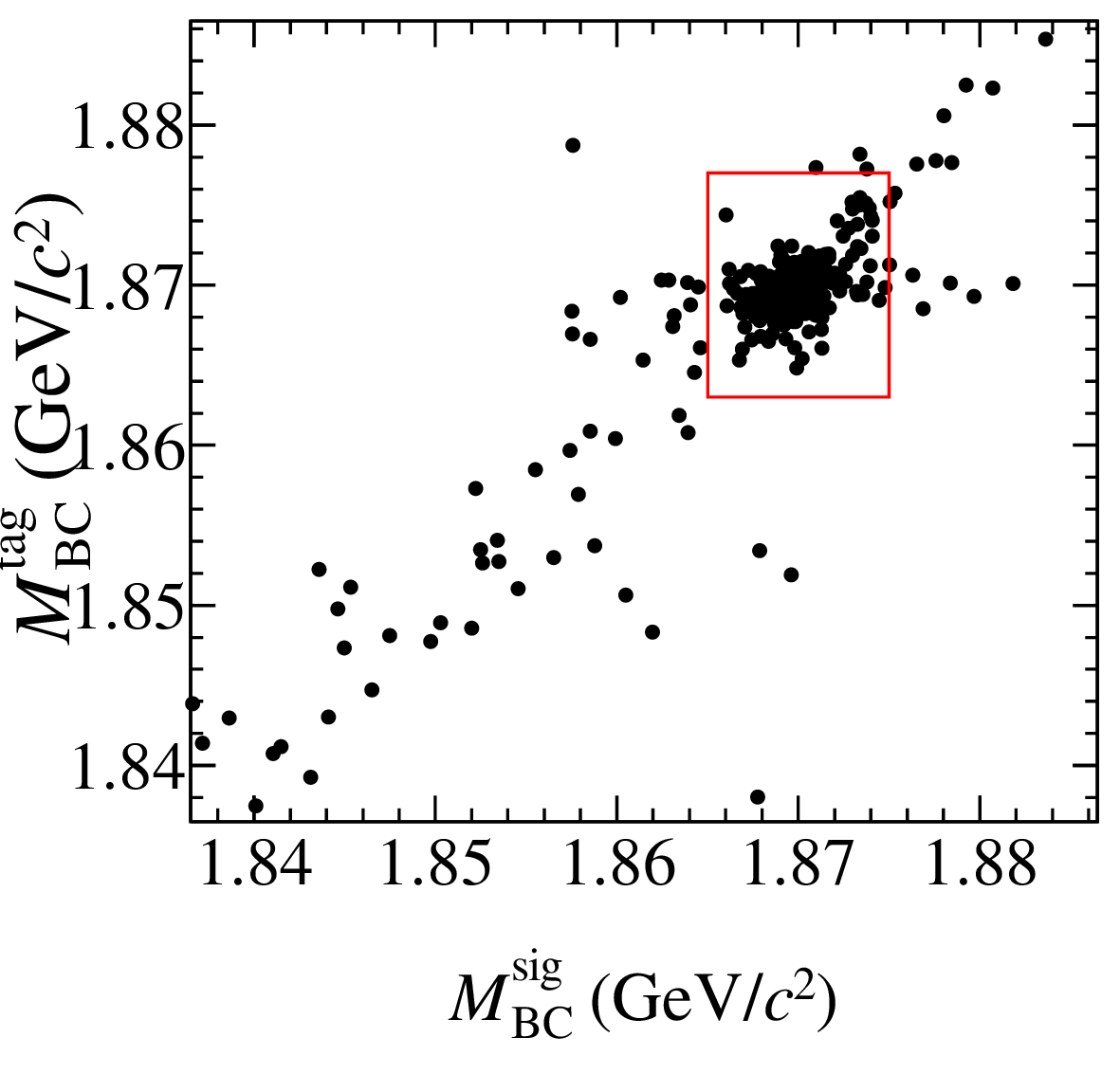}\\
  \includegraphics[width=5cm]{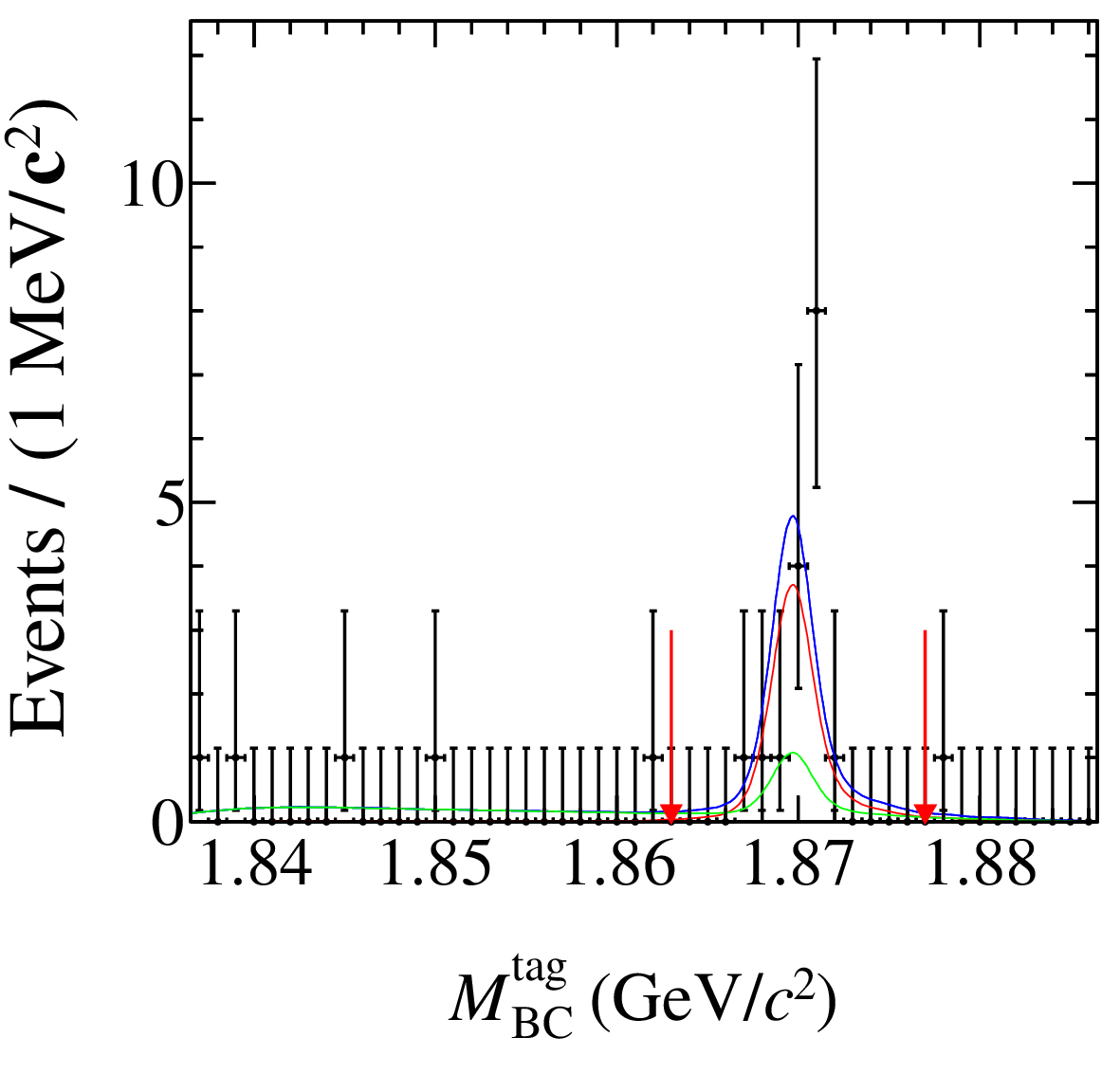}
  \includegraphics[width=5cm]{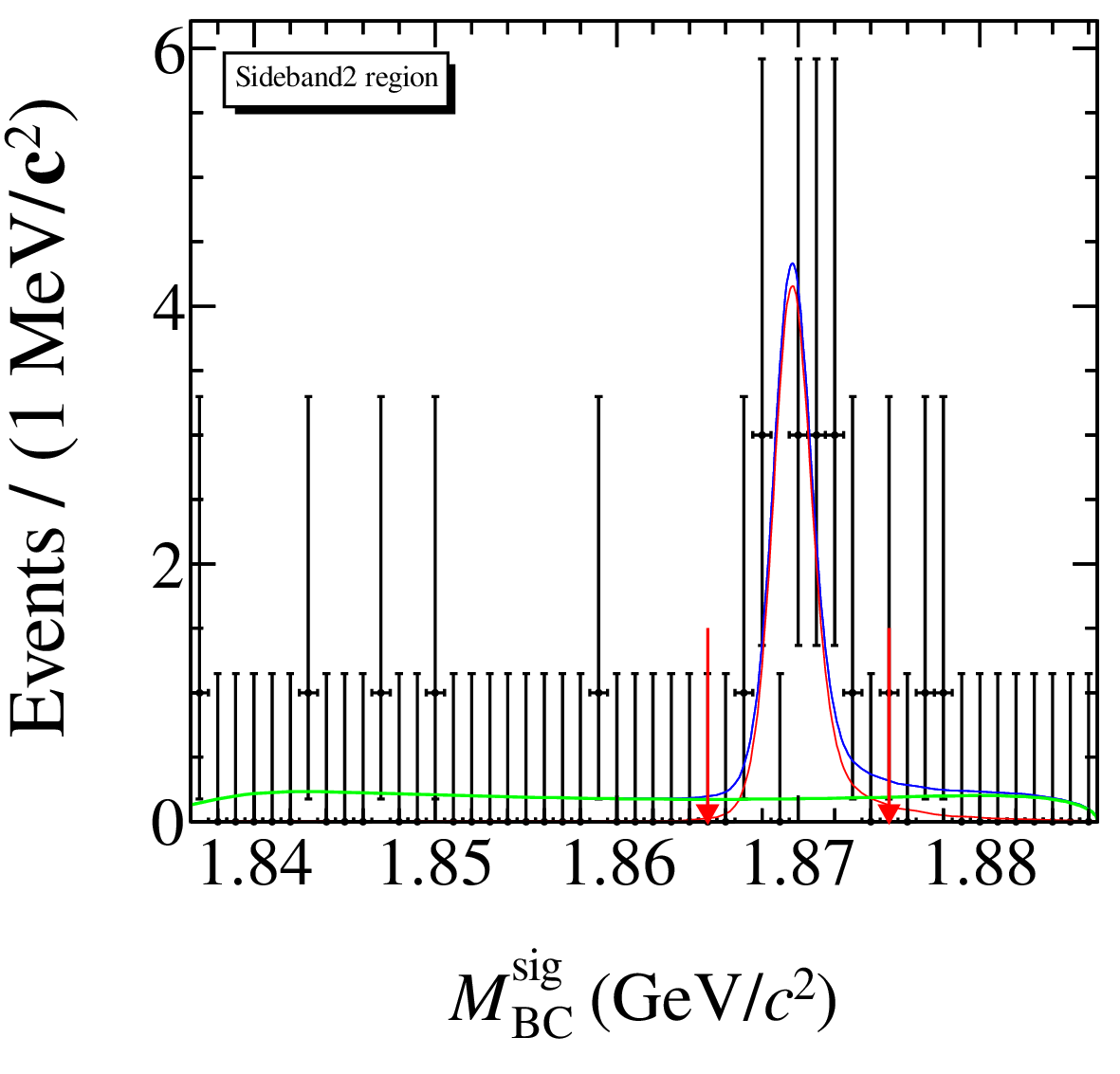}
  \includegraphics[width=5cm]{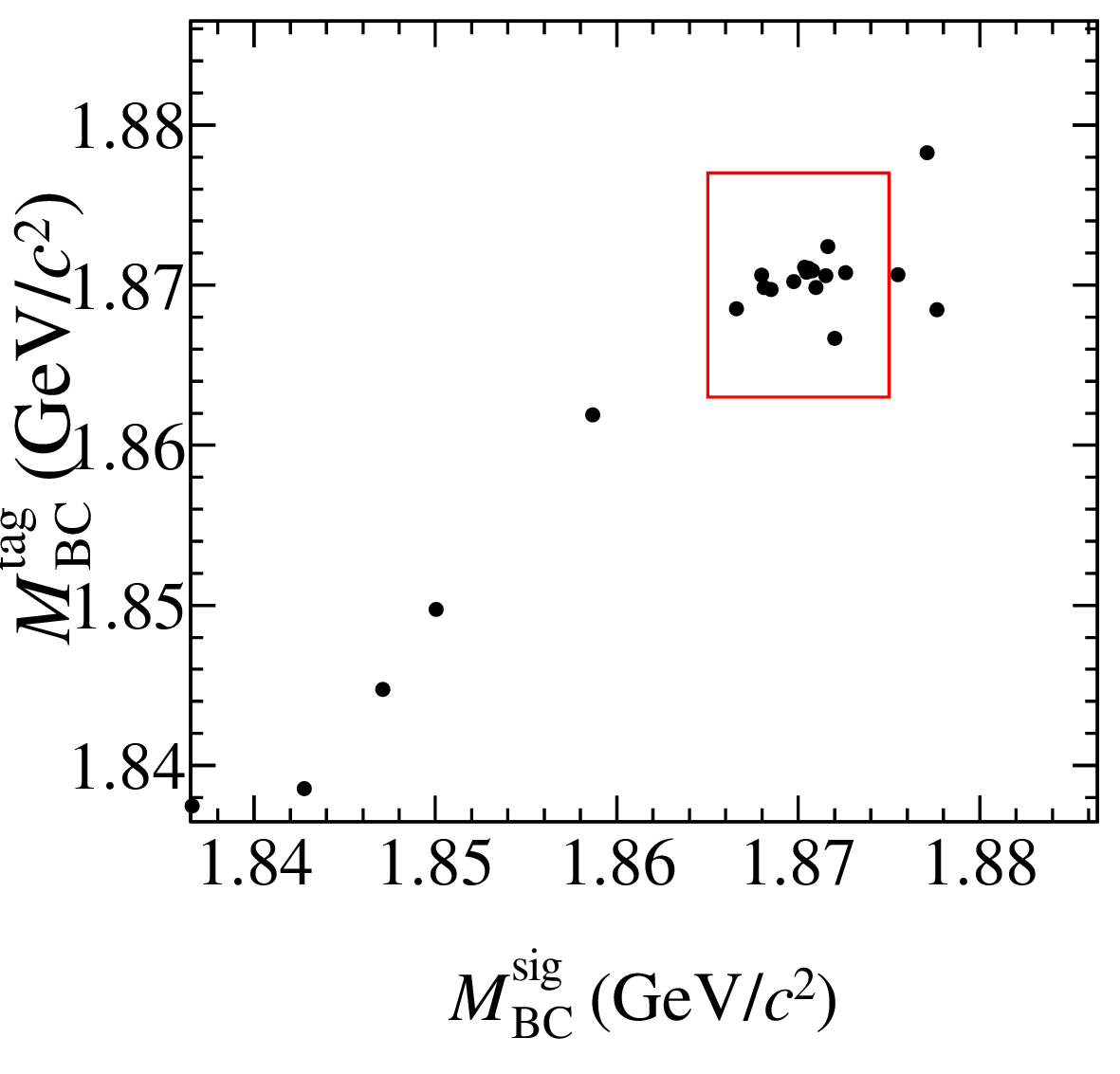}\\
   \caption{Projections on $M_{\rm BC}^{\rm tag}$ (left) and $M_{\rm BC}^{\rm sig}$ (center) of the 2D fit. The black points with error bars are data, the blue lines are the total fit, the red lines are the signal shapes, the green lines are the sum of all backgrounds, and the pair of red arrows indicate the $M_{\rm BC}$ signal window. The right plots are the $M_{\rm BC}^{\rm tag}$ versus $M_{\rm BC}^{\rm sig}$ distributions, where the red box represents the signal region of $M_{\rm BC}$. The first, second and third rows correspond to the fits to the candidate events in the signal region, the sideband 1 and the sideband 2 regions, respectively.}
  \label{fig:pwa_purity}
\end{figure*}

In this analysis, the combinatorial background in the $M_{\pi^{+}\pi^{-}}$ distribution is assumed to be flat,
and the net number of  $D^{+} \to K_{S}^{0}K_{S}^{0}\pi^{+}$ signal events is estimated as:
\begin{equation}
  \begin{aligned}
   N_{\rm net}=N_{K_S^0\rm sig}-\frac{1}{2}N_{\rm sb1}+\frac{1}{4}N_{\rm sb2},
  \end{aligned}	
  \label{eq:net}
\end{equation}
where $N_{K_S^0\rm sig}$, $N_{\rm sb1}$ and $N_{\rm sb2}$ are the numbers of $D^{+}$ meson in the signal, the sideband 1 and the sideband 2 regions, respectively,  from fitting the $M_{\rm BC}$ distributions of the accepted candidate events as explained in the next paragraph. The resulting $M_{\rm BC}$ distributions are shown in Fig.~\ref{fig:pwa_purity}, and the measured value of $N_{\rm net}$ is given in Table~\ref{tab:sidebandBF}.

\begin{figure}[htbp]
  \centering

  \includegraphics[width=7cm]{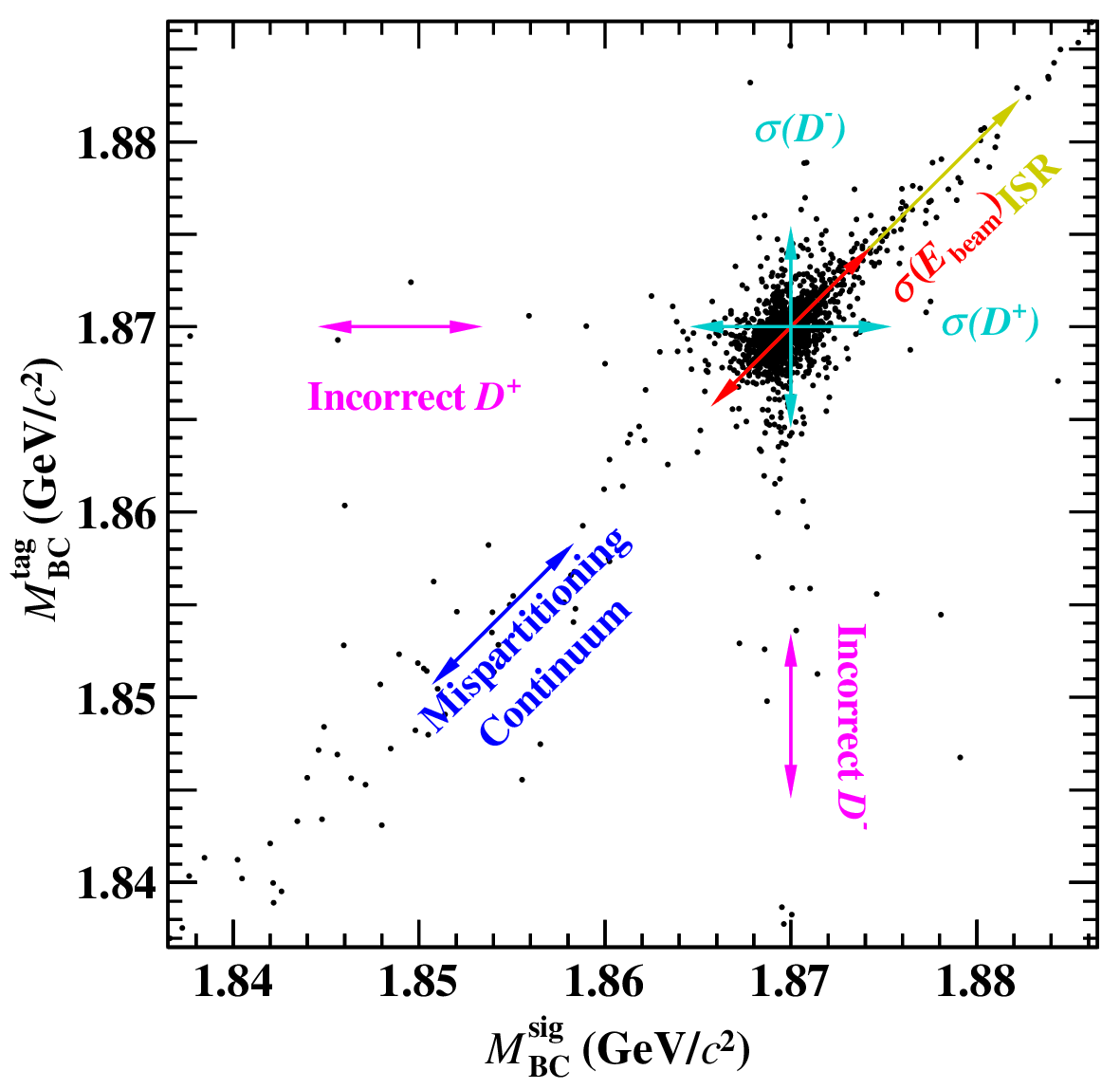} 
   \caption{Distribution of $M_{\rm BC}^{\rm tag}$ versus
    $M_{\rm BC}^{\rm sig}$ for the $D^{+}\to K_{S}^{0}K_{S}^{0}\pi^{+}$
    candidate events in data of $K_S^0$ signal region. The cyan arrows indicate the resolution caused by $D^{+(-)}$ momentum
    resolution~$(\sigma(D^{+(-)}))$, the red arrow by the beam energy spread~$(\sigma(E_{\rm beam}))$,  and the yellow arrow denotes the ISR effect. The pink and blue arrows indicate the background due to incorrectly reconstructed $D^-$ or $D^+$ and mispartitioning continuum, respectively. }
  \label{fig:2Dmbc}
\end{figure}

There are 1177 DT events obtained for the amplitude analysis with a signal
purity of $(89.1 \pm 2.9)\%$, which is determined from a 2D unbinned maximum
likelihood fit to the $M_{\rm BC}$ of the tag $D^{-}$ versus the $M_{\rm BC}$
of the signal candidates. Figure~\ref{fig:2Dmbc} shows the $M_{\rm BC}^{\rm tag}$ versus $M_{\rm BC}^{\rm sig}$ distribution for the
$D^{+} \to K_{S}^{0}K_{S}^{0}\pi^{+}$ candidate events in data.
The signal events with both the tag and the signal
side correctly reconstructed  concentrate around
$M_{\rm BC}^{\rm sig} = M_{\rm BC}^{\rm tag} = m_{D^+}$, where $m_{D^+}$ is
the known ${D^+}$ mass. The resolution of the signal peak is primarily influenced by the momentum resolution of the detector and the beam energy spread.   The $D^{+(-)}$ momentum resolution~$\sigma(D^{+(-)})$ causes the signal peak to be smeared along the
$M_{\rm BC}^{\rm sig}~(M_{\rm BC}^{\rm tag})$ axis, while the beam energy spread~$\sigma(E_{\rm beam})$
blurs the signal peak along the diagonal axis, $M_{\rm BC}^{\rm sig}+M_{\rm BC}^{\rm tag}$,
due to its correlation with the $M_{\rm BC}^{\rm sig}$ and $M_{\rm BC}^{\rm tag}$ variables.
Additionally, the signal peak is influenced by the ISR process,
which makes the available $E_{\rm beam}$ smaller than the effective beam energy,
leading to a diagonal ISR tail oriented toward the high diagonal axis.
Besides signal events, we define three kinds of background. Candidates with correctly
reconstructed $D^+$ (or $D^-$) and incorrectly reconstructed $D^-$ (or $D^+$) are
denoted as BG-I, which appear around the lines $M_{\rm BC}^{\rm sig}$ or
$M_{\rm BC}^{\rm tag} = m_{D^+}$. Other candidates appearing along the
diagonal (mispartitioning continuum) arise from
decay modes sharing the same final state particle combination with the signal~(BG-II).
In these cases, one or more particles are associated to the wrong $D$ meson, assuming all final state particles are well-reconstructed.
Consequently, the momentum of each signal or tag side is misreconstructed by the same amount,
resulting in a corresponding shift in the $M_{\rm BC}$ of each side by the same amount.
Moreover, some of the candidates smeared along the diagonal are from continuum events.
The remaining
background events mainly come from candidates reconstructed incorrectly on both sides (BG-III).
Here we list the PDFs for the corresponding components~\cite{2dfitPDFs,2Dfit-cleo} in the fit:

\begin{itemize}
    \item Signal: $s(x,y)$,
    \item BG-I: $b(x)\cdot {c_y}(y;E_{\rm beam}, \xi_y)+b(y)\cdot {c_x}(x;E_{\rm beam},\xi_x)$,
    \item BG-II: ${c_z}(z;\sqrt{2}E_{\rm beam},\xi_z) \cdot g(k; 0, \sigma_k)$,
    \item BG-III: ${c_x}(x;E_{\rm beam},\xi_x)\cdot {c_y}(y;E_{\rm beam},\xi_y)$.
\end{itemize}
Here, $x=M_{\rm BC}^{\rm tag}$, $y=M_{\rm BC}^{\rm sig}$, $z=(x+y)/\sqrt{2}$, and $k=(x-y)/\sqrt{2}$. The signal shape $s(x,y)$ is described by the 2D MC-simulated shape of $D^+D^-$ convolved with two independent Gaussian functions. For BG-I, the function $b(x)$ is described by the one-dimensional~(1D) MC-simulated shape of $D^+$($D^-$)  convolved with a Gaussian. The parameters of the Gaussian functions are obtained by a 1D fit to the $M_{\rm BC}$ distributions on the signal and tag side, and are fixed in the 2D fit.
The BG-II fit component is given by an ARGUS function in the diagonal axis multiplied by a Gaussian in the anti-diagonal axis. 
The function $c_{f}(f;E_{\rm beam},\xi_f)$ is the ARGUS function~\cite{argus} defined as
\begin{equation}
  \begin{aligned}
   c_{f}\left(f;E_{\rm beam},\xi_{f}\right)=A_{f}f\left(1-\frac{f^{2}}{E_{\rm beam}^{2}}\right)^{\frac{1}{2}}e^{\xi_{f}\left(1-\frac{f^2}{E_{\rm beam}^2}\right)},
  \end{aligned}	
  \label{eq:argus}
\end{equation}
where $f$ denotes $x,y$ or $z$, $E_{\rm beam}$ is fixed at 1.8865 GeV, $A_{f}$ is a normalization factor, and $\xi_f$ is a fit parameter. The function $g(k;\sigma_k)$ is a Gaussian distribution with a mean value of zero and a standard deviation $\sigma_k=\sigma_0 \cdot (\sqrt{2}E_{\rm beam}-z)^p$, where $\sigma_0$ and $p$ are parameters determined by the fit.
The BG-III fit component is given by an ARGUS function in $M_{\rm BC}^{\rm sig}$ multiplied by an ARGUS function in $M_{\rm BC}^{\rm tag}$. 
\begin{table}[hbtp]
\begin{center}
	\caption{Yields obtained from the 2D fit. The uncertainties are statistical only. }
	\begin{tabular}{ll cccc}
	\hline
	\hline
	&$N_{K_{S}^{0}\rm sig}$ &    1296 $\pm$ 37 
\\
	&$N_{\rm sb1}$ 	 & ~244 $\hspace{0.1em}\pm$ 16 
\\
	&$N_{\rm sb2}$   &~11 $\hspace{0.1em}\pm$ 4\\
\hline
	&$N_{\rm net}$   & 1177 $\pm$ 38\\

	\hline
	\hline
	\end{tabular}
	\label{tab:sidebandBF}
\end{center}
\end{table}

\section{V.~Amplitude analysis}
We perform a three-constraint kinematic fit to ensure that all the events lay within the phase space boundary, where the invariant masses of the signal $D^+$and the two $K_S^0$ candidates are constrained to the corresponding known masses~\cite{PDG}. The updated four-momenta of the final state particles from the kinematic fit are used to perform the amplitude analysis.
The amplitude analysis of $D^{+} \to K_{S}^{0}K_{S}^{0}\pi^{+}$ is performed using an unbinned maximum likelihood fit. The likelihood function $\mathcal{L}$ is constructed as
\begin{equation}
  \ln{\mathcal{L}} = \begin{matrix}\sum\limits_{k}^{N_{\rm data}} \ln [\omega_{\rm sig}f_{ S}(p_{j}^{k})+(1-\omega_{\rm sig})f_{B}(p^k_j)]\end{matrix},
  \label{eq:likefinal}
\end{equation}
where $k$ indicates the $k^{\rm th}$ event in the data sample, $N_{\rm data}$ is the number of surviving events, $p_j$ denotes the four-momenta of the $j^{\rm th}$ final state particles, $f_S$ ($f_B$) is the signal (background) PDF and $\omega_{\rm sig}$ is the signal purity of the DT events.

The signal PDF is given by
\begin{equation}
f_{S}(p_j) = \frac{\epsilon(p_j)|\mathcal M(p_j)|^2R_{3}(p_j)}{\int \epsilon(p_j)|\mathcal M(p_j)|^2R_{3}(p_j)\mathrm{d}p_j},
\label{pwa:pdf}
\end{equation}
where $\epsilon(p)$ is the detection efficiency and $R_3(p)$ is the three-body phase space. Based on the isobar model, the total amplitude $\mathcal{M}$ is constructed as the coherent sum of amplitudes of intermediate processes, given by $\mathcal M(p) = \sum{c_n\mathcal A_n(p)}$, where $c_n = \rho_ne^{i\phi_n}$ and $\mathcal A_n(p)$ are the complex coefficient and the amplitude for the $n^{\rm th}$ intermediate process, respectively. The magnitude $\rho_n$ and phase $\phi_n$ are free parameters in the fit. The model is symmetrized with respect  to the two identical $K_S^0$ mesons.  We use covariant tensors to construct the amplitudes, which are written as
\begin{equation}
  \mathcal{A}_{n} = P_{n}S_{n}F_{n}^{r}F_{n}^{D},
\end{equation}
where $S_{n}$ is the spin factor~\cite{covariant-tensors}, $F_{n}^{r}$ and $F_{n}^{D}$ are the Blatt-Weisskopf barrier factors of the intermediate state and the $D^{\pm}$ meson decays~\cite{Blatt}, respectively, and $P_{n}$ is the relativistic Breit-Wigner (RBW) function~\cite{RBW} describing the propagator of the intermediate resonance.
The background PDF is given by
\begin{equation}
f_B(p_j)=\frac{B(p_j)R_3(p_j)}{\int{\epsilon(p_j)B_{\epsilon}(p_j)R_3(p_j)}\mathrm{d}p_j},
\label{bkglikelihood}
\end{equation}
where $B_{\epsilon}(p_j)=B(p_j)/\epsilon(p_j)$ is the efficiency-corrected background shape, the $\epsilon(p_j)$ term is given by the number of selected events from the PHSP MC samples divided by the number of generated events in the $M^{2}_{k_S^0K_S^0}$ versus $M^{2}_{K_S^0\pi^+}$ 2D space. The shape of the background in data is modeled by the background events in the signal region derived from the inclusive MC samples. The invariant mass distributions of events outside the signal region show good agreement between data and MC simulation, thus validating the description from the inclusive MC samples. We have also examined the distributions of the background events of the inclusive MC samples inside and outside the signal region. Generally, they are consistent with each other within statistical uncertainties.  The background shape $B(p)$ is modeled using a kernel estimation method~\cite{CRANMER2001198} implemented in RooFit~\cite{RooNDKeysPDF} to model the distribution of an input dataset as a superposition of Gaussian kernels. The normalization integrals in the denominators are calculated by the phase space MC sample, which are 
\begin{equation}
\int \epsilon|\mathcal M(p_j)|^{2}R_{3} \mathrm{d}p_{j} \propto \frac{1}{N_{\rm MC}}\sum^{N_{\rm MC}}_{k_{\rm MC}}\frac{|\mathcal M(p^{k_{\rm MC}}_j)|^2}{|\mathcal M^{\rm gen}(p^{k_{\rm MC}}_j)|^2},
  \label{integral}
\end{equation}
\begin{equation}
    \int \epsilon B_{\epsilon}(p_{j}) R_{3} \mathrm{d}p_{j} \propto \frac{1}{N_{\rm MC}} \sum^{N_{\rm MC}}_{k_{\rm MC}}\frac{B_{\epsilon} (p_{j}^{k_{\rm MC}}) }{|\mathcal M^{\rm gen}(p_{j}^{k_{\rm MC}})|^2},
\end{equation}
where $k_{\rm MC}$ is the index of the $k^{\rm th}$ event of the MC sample and $N_{\rm MC}$ is the number of selected MC events. The $\mathcal M^{\rm gen}(p_j)$ is the signal PDF used to generate the MC samples in MC integration.

Tracking, PID and $K_{S}^{0}$ reconstruction efficiency differences between data and MC simulation are corrected by multiplying the MC events by a correction factor $\gamma_{\epsilon}$, which is calculated as
\begin{equation}
  \gamma_{\epsilon}(p_j) = \prod_{n} \frac{\epsilon_{n,\rm data}(p_j)}{\epsilon_{n,\rm MC}(p_j)},
  \label{pwa:gamma}
\end{equation}
where $n$ refers to tracking, PID or $K_S^0$ reconstruction, and $\epsilon_{n,\rm data}(p_j)$ and $\epsilon_{n,\rm MC}(p_j)$ are their  efficiencies as a function of the momenta of the daughter particles for data and MC simulation, respectively. The specific values of these efficiencies are obtained using different control samples. More detailed information can be found in section 7, as part of the uncertainty studies for the BF measurement. By weighting each MC event with $\gamma_{\epsilon}$, the MC integration is modified to be
\begin{equation}
\int \epsilon|\mathcal M(p_j)|^{2}R_{3}\mathrm{d}p_{j} \propto \frac{1}{N_{\rm MC}}\sum^{N_{\rm MC}}_{k_{\rm MC}}\frac{\gamma_{\epsilon}|\mathcal M(p^{k_{\rm MC}}_j)|^2}{|\mathcal M^{\rm gen}(p^{k_{\rm MC}}_j)|^2},
\end{equation}
\begin{equation}
    \int \epsilon B_{\epsilon} R_{3}\mathrm{d}p_{j} \propto \frac{1}{N_{\rm MC}} \sum^{N_{\rm MC}}_{k_{\rm MC}}\frac{\gamma_{\epsilon}B_{\epsilon} (p_{j}^{k_{\rm MC}}) }{|\mathcal M^{\rm gen}(p_{j}^{k_{\rm MC}})|^2}.
\end{equation}


For a decay process $a \to b\ c$, the Blatt-Weisskopf barrier {{factors}}~\cite{Blatt} depend on the angular momentum $L$ and the momentum $q$ of the final-state particle $b$ or $c$ in the rest system of $a$. They are taken as
\begin{equation}
\begin{aligned}
  &F_{L=0}(q)=1,\\
  &F_{L=1}(q)=\sqrt{\frac{z_0^2+1}{z^2+1}},\\
  &F_{L=2}(q)=\sqrt{\frac{z_0^4+3z_0^2+9}{z^4+3z^2+9}}, \label{xl}
\end{aligned}
\end{equation}
where $z = qR$ and $z_0 = q_0R$. The effective radius of barrier $R$ is fixed to be 3.0 $(\rm {GeV}/c)^{-1}$ for the intermediate resonances and 5.0 $(\rm {GeV}/c)^{-1}$ for the $D^+$ meson. The momentum $q$ is given by
\begin{equation}
q = \sqrt{\frac{(s_a+s_b-s_c)^2}{4s_a}-s_b}, \label{q2}
\end{equation}
where $s_a, s_b,$ and $s_c$ are the invariant mass squared of particles $a,\ b$ and $c$, respectively. The value of $q_0$ is that of $q$ when $s_a = m^2_a$, where $m_a$ is the mass of particle $a$.

The intermediate resonances $K^{*}(892)^+ ,K_{2}^{*}(1430)^{+}$, $f_{0}(1370)$, $a_{0}(1450)^{0}$, $K_{0}^{*}(1680)^{+}$ are parameterized with the RBW formulas,
\begin{equation}
\begin{aligned}
		&P(m) = \frac{1}{m^2_0-m^2-im_0\Gamma(m)}, \\ 
		&\Gamma(m)=\Gamma_0\left(\frac{q}{q_0}\right)^{2L+1}\left(\frac{m_0}{m}\right)\left(\frac{F_L(q)}{F_L(q_0)}\right)^2,  \label{propagator}
\end{aligned}
\end{equation}
where $m$ is the invariant mass of the decay products, and $m_0$ and $\Gamma_0$ are the rest mass and width of the intermediate resonance, which are fixed to their known values~\cite{PDG}. The energy-dependent width is denoted by $\Gamma(m)$.

The $K\pi$ $\mathcal{S}$-wave modeled by the LASS parameterization~\cite{Aston:1987ir} is described by a $K^*_0(1430)$ Breit-Wigner together with an effective range non-resonant component with a phase shift, given by
\begin{equation}
A(m) = F{\rm sin}\delta_F e^{i\delta_F} + R{\rm sin}\delta_R e^{i\delta_R}e^{i2\delta_F},
\end{equation}
with
\begin{equation}
\begin{aligned}
&\delta_F = \phi_F + {\rm cot}^{-1}\left[\frac{1}{aq}+\frac{rq}{2}\right],\\
&\delta_R = \phi_R + {\rm tan}^{-1}\left[\frac{M\Gamma(m_{K\pi})}{M^2-m^2_{K\pi}}\right],
\end{aligned}
\end{equation}
where the parameters $F~(\phi_F)$ and $R~(\phi_r)$ are the magnitudes (phases) for the non-resonant state and resonance terms, respectively. The parameters $a$ and $r$ are the scattering length and effective interaction length, respectively.  $M$ and $\Gamma$ are the BW parameters of the $K^*_0(1430)$. We fix these parameters ($M, \Gamma, F, \phi_F, R, \phi_R, a, r$) to the  \mbox{\slshape B\kern-0.1em{\smaller A}\kern-0.1em B\kern-0.1em{\smaller A\kern-0.2em R}} and Belle results~\cite{KpiLASS}. These parameters are summarized in Table~\ref{LASSPa}.

\begin{table}[hbtp]
    \centering
    \caption{The $K\pi$ $\mathcal{S}$-wave parameters, obtained from the fit to the $D^0 \to K_S^0 \pi^{+} \pi^{-}$ Dalitz plot distribution in the \mbox{\slshape B\kern-0.1em{\smaller A}\kern-0.1em B\kern-0.1em{\smaller A\kern-0.2em R}} and Belle experiments~\cite{KpiLASS}. Uncertainties are statistical only.}
    \begin{tabular}{l ccc}
    \hline
    \hline
      &  $M\ (\rm GeV/ \textit c^{2})$ \quad & $1.441 \pm 0.002$  \\
      &  $\Gamma\ (\rm GeV$) \quad & $0.193 \pm 0.004$ \\    &  $F$ \quad & $0.96 \pm 0.07$ \\
      &  $\phi_{F}\ (^\circ)$ \quad & $0.1 \pm 0.3$ \\
      &  $R$ \quad & 1(fixed) \\
      &  $\phi_{R}\ (^\circ)$ \quad & $\hspace{-1.7em}-109.7 \pm 2.6$ \\
      &  $a\ (\rm GeV/ \textit c^{2})$ \quad & $0.113 \pm 0.006$ \\
      &  $r\ (\rm GeV/ \textit c^{2})$ \quad & $\hspace{-1.2em}-33.8 \pm 1.8$ \\
  \hline
  \hline
    \end{tabular}
    \label{LASSPa}
\end{table}

The $f_0(980)$ is parameterized with the Flatt\'e formula~\cite{f0980}:
\begin{equation}
	P_{f_0(980)}=\frac{1}{M_{f_0(980)}^2-m^2-i(g_{\pi\pi}\rho_{\pi\pi}(m^2)+g_{K\bar{K}}\rho_{K\bar{K}}(m^2))},
\end{equation}
where $g_{\pi\pi,K\bar{K}}$ are the coupling constants to individual final states. The parameters are fixed to be $g_{\pi\pi}=(0.165{ \pm 0.010 \pm 0.015)} {\rm GeV^2}/c^4$, $g_{K\bar{K}}=(4.21{ \pm 0.25 \pm 0.21)}g_{\pi\pi}$ and $M = 0.965$~GeV/$c^2$, as reported in Ref.~\cite{f0980}. The Lorentz invariant phase space factors $\rho_{\pi\pi}(s)$ and $\rho_{K\bar{K}}(s)$ are given by
\begin{equation}
\begin{aligned}
	\rho_{\pi\pi}&=\frac{2}{3}\sqrt{1-\frac{4m^2_{\pi^\pm}}{m^2}}+\frac{1}{3}\sqrt{1-\frac{4m^2_{\pi^0}}{m^2}},\\
	\rho_{K\bar{K}}&=\frac{1}{2}\sqrt{1-\frac{4m^2_{K^\pm}}{m^2}}+\frac{1}{2}\sqrt{1-\frac{4m^2_{K^0}}{m^2}}.
\end{aligned}
\end{equation}

\begin{figure}[htbp]
  \centering
  \includegraphics[width=0.235\textwidth,trim=0 0 2cm 0,clip]{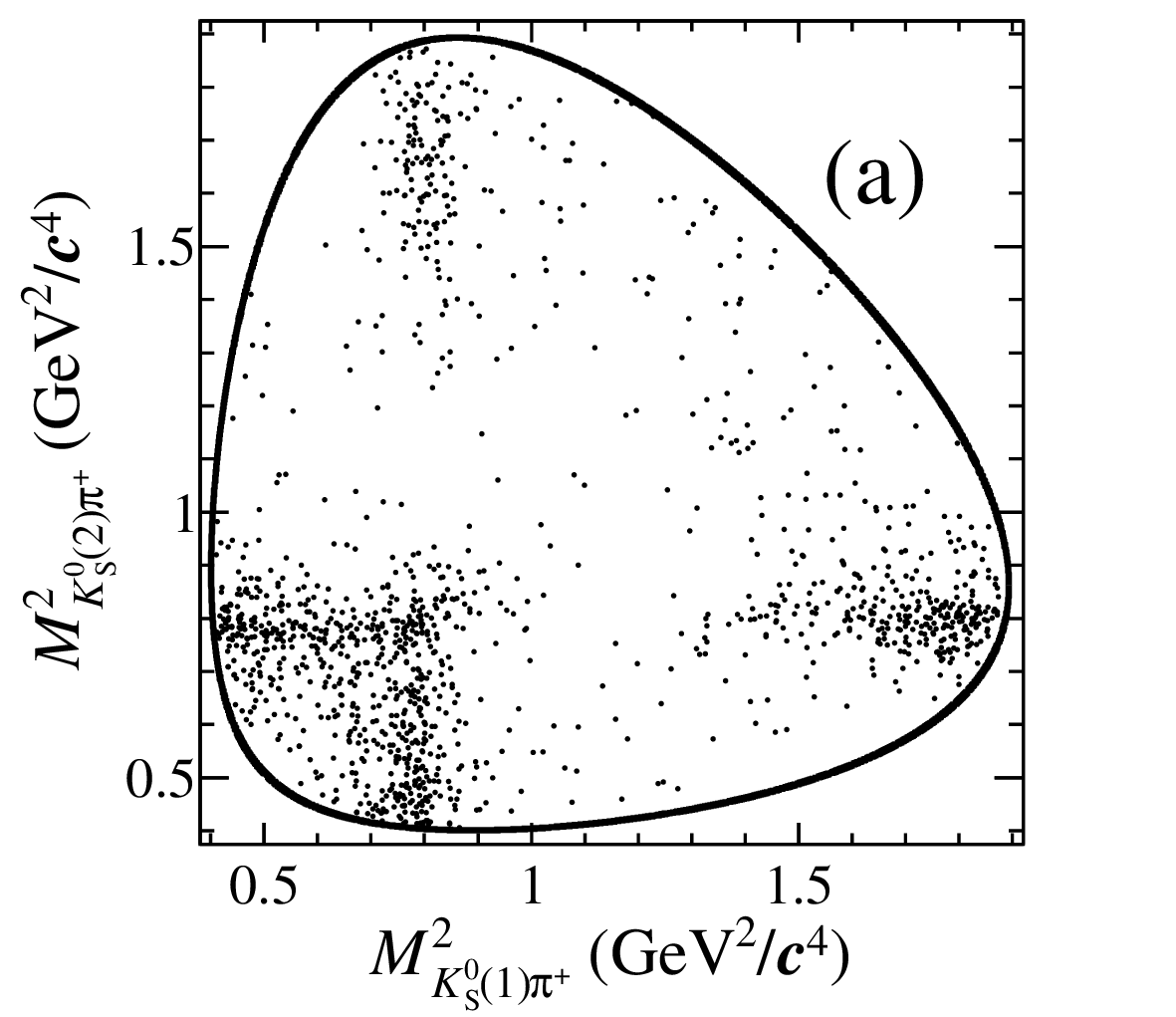}
  \includegraphics[width=0.235\textwidth,trim=0 0 2cm 0,clip]{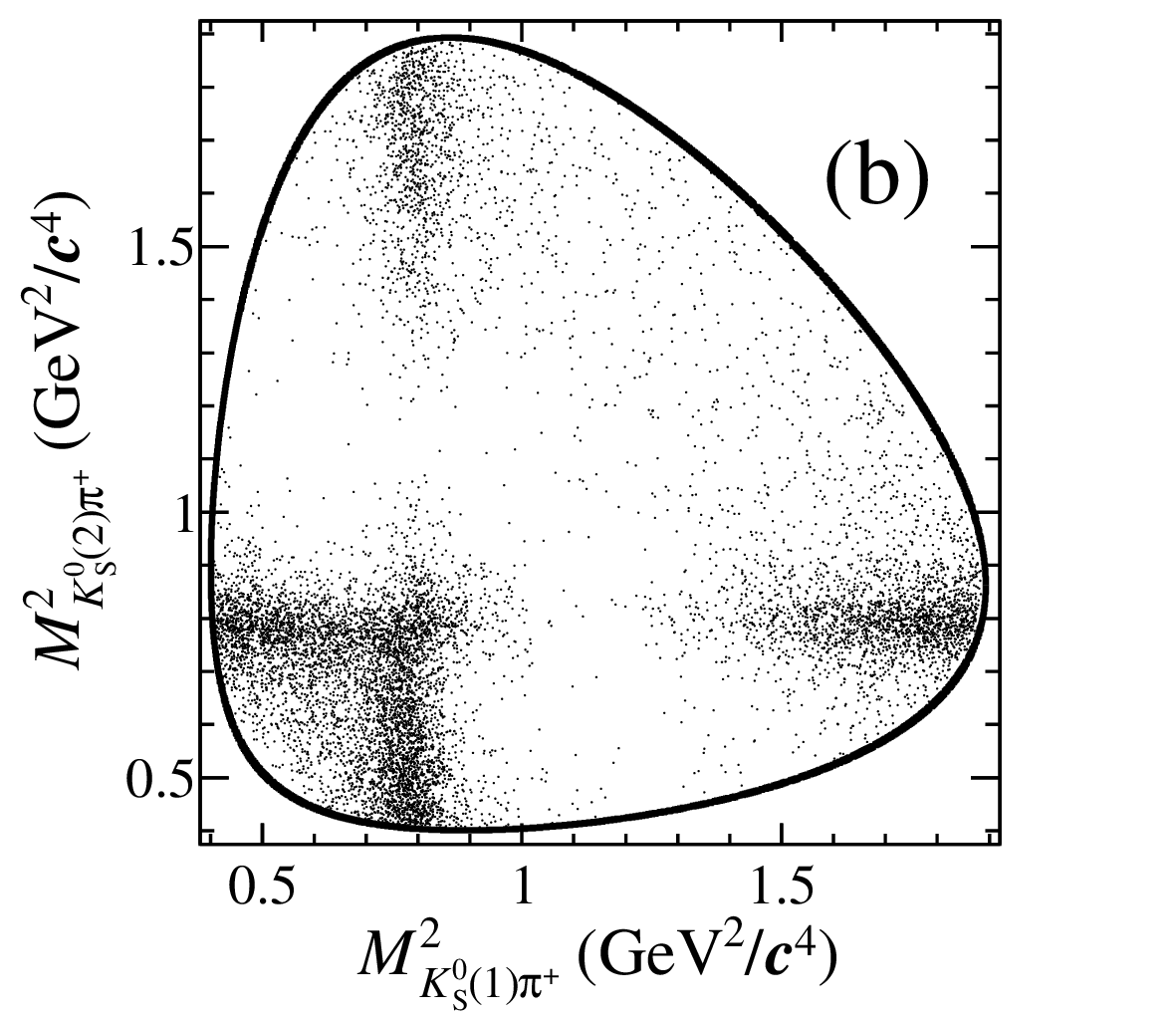}
    \caption{Dalitz plots of $M^{2}_{K_S^0(1)\pi^+}$ versus~$M^{2}_{K_S^0(2)\pi^+}$
      for $D^+\to K^0_SK^0_S\pi^+$, $K_{S}^0(1)$ and $K_{S}^0(2)$ are randomly selected from two indistinguishable $K_S^0$ candidates, of (a) the sum of all data samples and (b) the signal MC samples generated
      based on the amplitude analysis result. The black curves indicate the
      kinematic boundary.}
    \label{fig:dalitz}
\end{figure}

\begin{figure}[!htbp]
  \centering
  \hspace{-0.em}\includegraphics[width=0.455\textwidth]{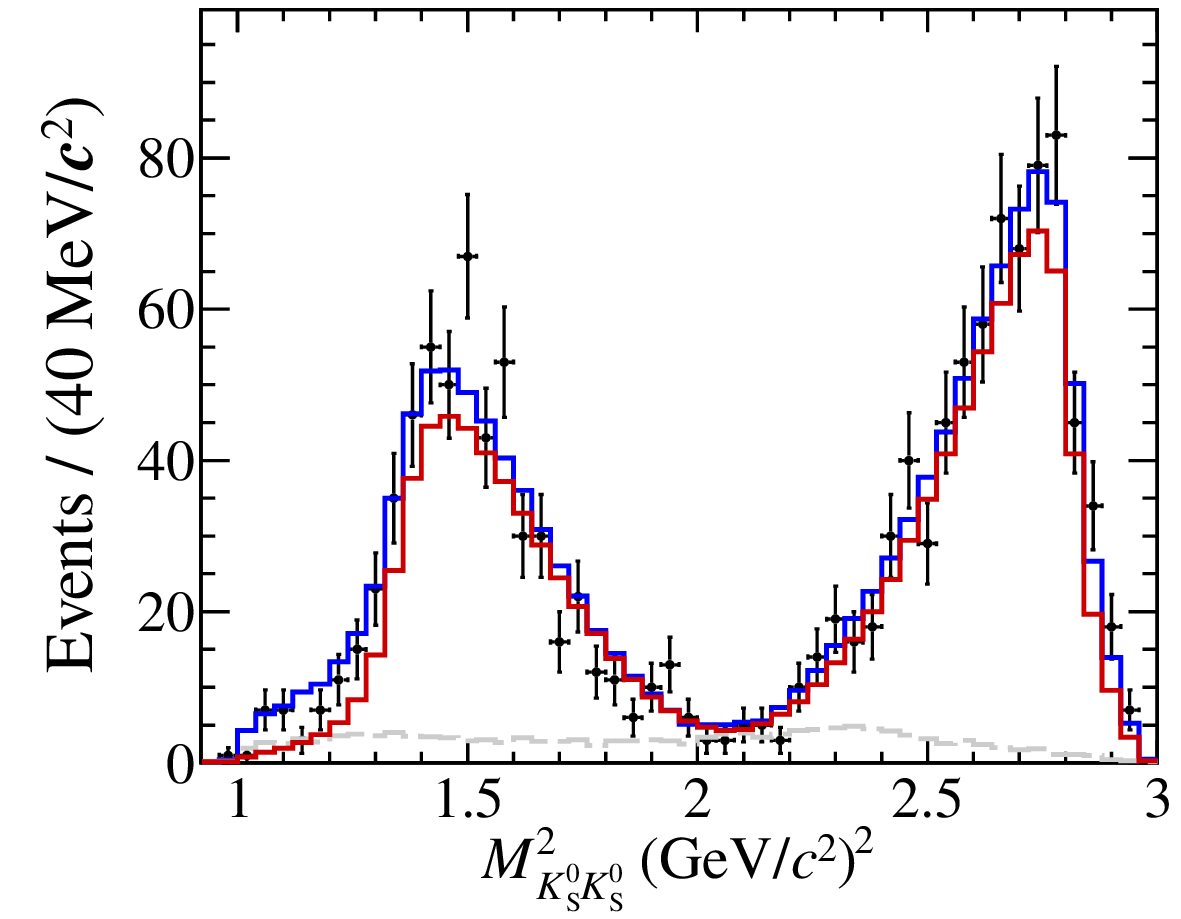}\\
  \hspace{-0.em}\includegraphics[width=0.455\textwidth]{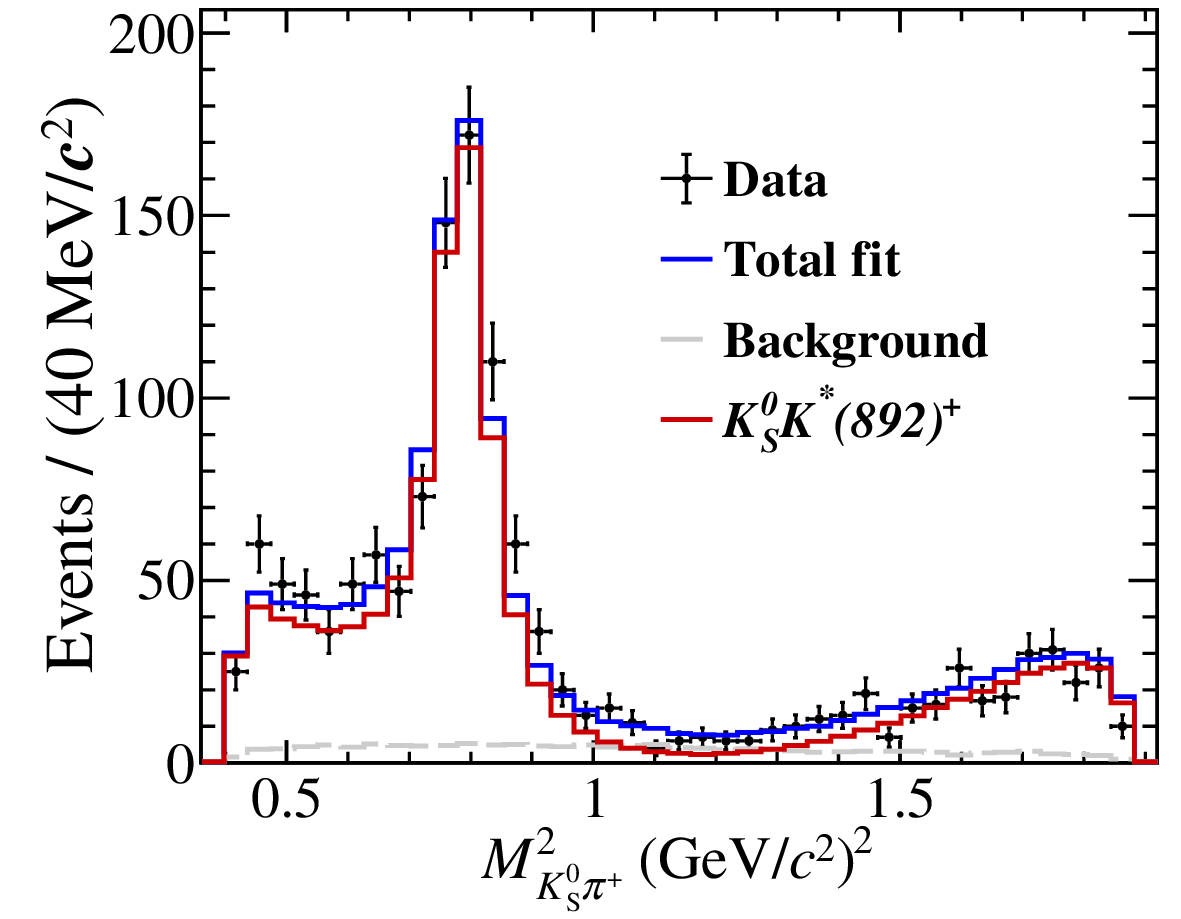}
  \caption{Distributions of  $M^{2}_{K_S^0K_S^0}$ and  $M^{2}_{K_S^0\pi^+}$ from the nominal fit. The $M^{2}_{K_{S}^{0}\pi^{+}}$ distribution only contains one $K_{S}^{0}\pi^{+}$ combination, in which $K_{S}^{0}$ is labelled randomly due to the indistinguishable $K_{S}^{0}$ candidates. The points with error bars are data and the blue lines are the total fit result. The red solid lines are the $D^{+}\to K_{S}^{0}K^{*}(892)^{+}$ component. The dashed grey lines are the simulated background derived from the inclusive MC sample.}
  \label{dalitz-projection}
\end{figure}

Figure~\ref{fig:dalitz}(a) shows the Dalitz plot of $M^2_{K_S^0(1)\pi^+}$ versus $M^2_{K_S^0(2)\pi^+}$ of the selected DT candidates from the data samples, where $K_{S}^0(1)$ and $K_{S}^0(2)$ are randomly selected from two indistinguishable $K_S^0$ candidates. The vertical and horizontal bands around 0.8~GeV$^2$/$c^4$ indicate the decay $D^+\to K_S^0K^{*}(892)^+$.
We choose this process as a reference so that  magnitudes and phases of other amplitudes are defined relative to it. Other possible contributions from the resonances such as  $f_{0}(980)$, $f_{0}(1370)$, $K_{2}^{*}(1430)^{+}$, $a_{0}(1450)^{0}$, $K_{0}^{*}(1680)^{+}$ and $K\pi$ $\mathcal{S}$-wave  are added to the fit one at a time.  The statistical significance of each resonance is calculated from the change of the log-likelihood taking the variation in the number of degrees of freedom into account. Different combinations of these resonances
are also tested. In addition to the reference amplitude $D^+\to K_S^0K^{*}(892)^+$,
the amplitude for  $D^+\to K_{S}^{0}(K_{S}^{0}\pi^{+})_{ \mathcal{S}-\rm wave}$ is found to have a significance larger than $10\sigma$.
No other contribution has a significance larger than $3\sigma$.
The significance of the $f_{0}(980)$ contribution is less than $1\sigma$.
The Dalitz plot of the signal MC sample
generated based on the result of the amplitude analysis is shown in Fig.~\ref{fig:dalitz}(b).
The mass projections of the fit are shown in Fig.~\ref{dalitz-projection}. The fit quality is determined as $\chi^2/{\rm NDF = 113.1/125}$ using an adaptive binning of the $M_{K_S^0K_S^0}^{2}$ versus $M_{K_{S}^{0}\pi^{+}}$, where the Dalitz plot that each bin requires must contain at least 10 events.

The contribution of the $n^{\rm th}$ amplitude relative to the total BF is quantified
by the fit fraction~(FF) defined as
${\rm FF}_{n} = \int \left|\rho_{n}A_{n}\right|^{2}dR_3/\int\left|\mathcal M\right|^{2}dR_3$.
The FFs  and the phase difference relative to the reference process are listed in Table~\ref{fit-result}.
The sum of the two FFs is  102.2$\%$. The sum
of these FFs is generally not unity due to net constructive or destructive interference.
\begin{table}[!htbp]
  \caption{The obtained FFs and phase difference to the reference amplitude. The first and the second uncertainties are statistical and systematic, respectively. }
  \label{fit-result}
  \begin{center}
    \begin{tabular}{lccc}
      \hline \hline
      Amplitude                                 & Phase                            & FF~(\%)                \\
      \hline
      $D^{+} \to K_{S}^{0}K^{*}(892)^{+}$   & 0.0(fixed)                       & $97.8 \pm 1.0 \pm 0.3$  \\
      $D^{+} \to K_{S}^{0}(K_{S}^{0}\pi^{+})_{\mathcal{S}-\rm wave}$        & \phantom{0}$1.4 \pm 0.2 \pm 0.1$ & $\hspace{0.5em}4.4 \pm 1.0 \pm 0.2$  \\
      \hline \hline
    \end{tabular}
  \end{center}
\end{table}
\subsection{VI.~SYSTEMATIC UNCERTAINTY OF THE AMPLITUDE ANALYSIS}
\label{sec:syss1}
The systematic uncertainties for the amplitude analysis results, including the
phase and FFs, can be categorized into the following sources: (I) input parameters, (II) radii of the mesons, (III) background, (IV) efficiency, (V) fit bias. For each kind of source, the related parameters are varied and the results of the new fit are compared to the results of the nominal fit, considering the difference as the corresponding uncertainty.
The results of the systematic uncertainties for phase and FFs are summarized in Table~\ref{ampsys}, where the uncertainties are given in units of the corresponding statistical uncertainties. The total uncertainties are obtained by adding these contributions in quadrature.

\begin{table}[htbp]
 \caption{Systematic uncertainties on the phase $\phi$ and FFs for different amplitudes in units of the corresponding statistical uncertainties. (I) input parameters, (II) radii of the mesons, (III) background, (IV) efficiency, (V) fit bias. The total systematic uncertainties are obtained by summing up all contributions in quadrature.}
  \centering
\begin{tabular}{lcccc}
\hline \hline & $D^{+} \rightarrow K_S^0 K^*(892)^{+} $ & & \multicolumn{2}{c}{$D^{+} \to K_{S}^{0}(K_{S}^{0}\pi^{+})_{\mathcal{S}-\rm wave}$} \\
\cline { 2 - 2 } \cline { 4 - 5 } Source & FF & & Phase $\phi$ & FF \\
\hline 
I     & 0.12 & & 0.15 & 0.06 \\
II    & 0.12 & & 0.14 & 0.06 \\
III   & 0.13 & & 0.12 & 0.10 \\
IV    & 0.04 & & 0.40 & 0.13 \\
V     & 0.07 & & 0.06 & 0.08 \\
Total & 0.23 & & 0.47 & 0.20 \\
\hline \hline
\label{ampsys}
\end{tabular}
\end{table}
(I) {\textit{Input parameters}}: in the nominal fit,  mass and width of the $K^{*}(892)^{+}$ are fixed to the values from PDG~\cite{PDG}. The parameters of the LASS model are fixed according to Ref.~\cite{KpiLASS}. To estimate the corresponding systematic uncertainties, the fit procedure is repeated by varying one by one each of the fixed parameters by $\pm 1 \sigma$. The quadratic sum of the largest relative variations for each parameter is taken as the systematic uncertainty.

(II) {\textit{Radii of the mesons}:}  the radii of the Blatt-Weisskopf barrier factors are varied from their nominal values of $5$~GeV$^{-1}$ and $3$~GeV$^{-1}$ (for the $D^{+}$ meson
and the intermediate resonances, respectively)  by $\pm 1$~GeV$^{-1}$.

(III) {\textit{Background}:} the uncertainties from the background size is studied by varying the signal fraction (equivalent to the fraction of background), i.e. $\omega_{\rm sig}$ in Eq.~(\ref{eq:likefinal}), by $\pm 1\sigma$. Another source of systematic uncertainty comes from the modelization of the background shape, which is estimated by varying the parameter of RooNDKeysPDF describing the smoothness of the shape from 1.5 to 2 or 1, and extracting the shape with  two different variables ($M_{K_S^0K_S^0}^{2}$ and $M_{K_{S}^0\pi^+}^{2}$). The largest difference from the nominal results is considered as the uncertainty.

(IV) {\textit{Efficiency}:} the systematic uncertainty from the $\gamma_{\epsilon}$ factor in Eq.~(\ref{pwa:gamma}), which corrects for data-MC differences in tracking, PID and $K_S^0$ reconstruction efficiencies, is evaluated by performing the fit after varying the $\gamma_{\epsilon}$ factor by $\pm 1\sigma$. 

Tracking and PID efficiencies of $\pi^{+}$  are studied by the control samples of $D^{0}\to K^{-}\pi^{+}$, $D^{0}\to K^{-}\pi^{+}\pi^{0}$, $D^{0}\to K^{-}\pi^{+}\pi^{+}\pi^{-}$ versus $\bar{D}^{0}\to K^{+}\pi^{-}$, $\bar{D}^{0}\to K^{+}\pi^{-}\pi^{0}$, $\bar{D}^{0}\to K^{+}\pi^{-}\pi^{-}\pi^{+}$ and $D^{+} \to K^{-}\pi^{+}\pi^{+}$ versus $D^{-} \to K^{+}\pi^{-}\pi^{-}$. The $K_{S}^{0}$ reconstruction efficiency is studied by the control samples from $\psi(3770) \to D^{0}\bar{D}^{0}/D^{+}D^{-}$ and selected by DT method. The tag modes are $\bar{D}^{0} \to K^{-}\pi^{+},~ \bar{D}^{0} \to K^{-}\pi^{+}\pi^{0},~ \bar{D}^{0} \to K^{-}\pi^{+}\pi^{+}\pi^{-}$,~ $D^{-} \to K^{+}\pi^{-}\pi^{-}$ and the signal modes are: $D^{0} \to K_{S}^{0}\pi^{+}\pi^{-},~D^{0} \to K_{S}^{0}\pi^{+}\pi^{-}\pi^{0},~ D^{0} \to K_{S}^{0}\pi^{0},~D^{+} \to K_{S}^{0}\pi^{+},~D^{+} \to K_{S}^{0}\pi^{+}\pi^{0},~D^{+} \to K_{S}^{0}\pi^{+}\pi^{+}\pi^{-}$.

(V) {\textit{Fit bias}:} to study the possible bias from the fit procedure, an ensemble of 300 signal MC samples is generated according to the results of the amplitude analysis. The fit procedure is repeated for each signal MC sample, and the pull distributions of the amplitude results are fitted by a Gaussian. The obtained  mean values are assigned as the correlated  systematic uncertainty. 

\section{VII.~Branching fraction measurement}
\begin{figure}[htp]
  \begin{center}
    \includegraphics[width=0.48\textwidth]{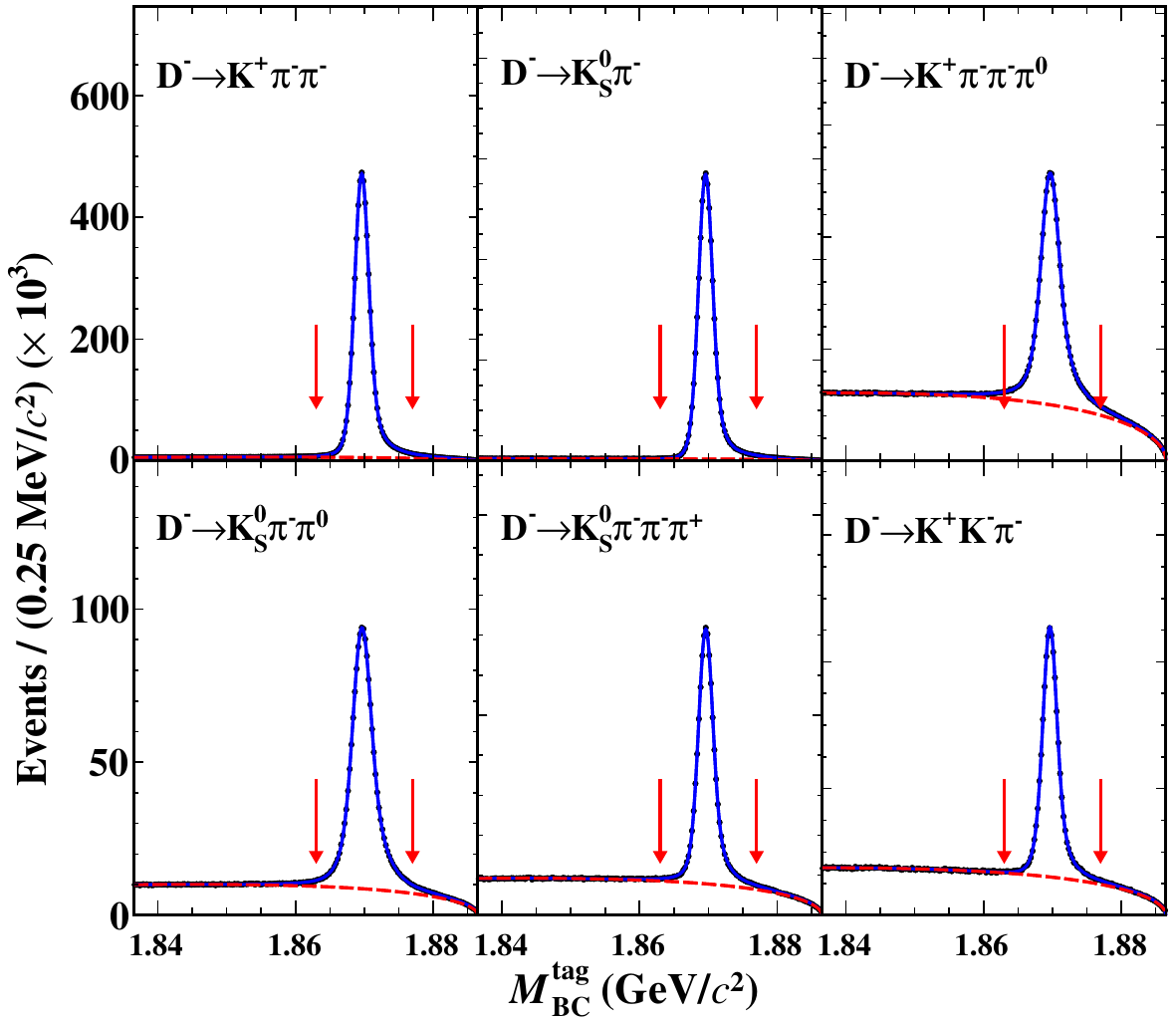}
    \caption{Fits to the $M_{\rm BC}^{\rm tag}$  distributions of the ST $D^{-}$
      candidates. The data samples are represented by points with error bars,
      the total fit results by solid blue lines and the red dashed curves describe the fitted combinatorial background shapes. The pair of red arrows indicate the $M_{\rm BC}$ signal window.}
    \label{fig:DT_fit}
  \end{center}
\end{figure}

The BF of $D^+\to K_S^0K_S^0\pi^+$ is measured with the DT technique
using the same tag modes and event selection criteria as  the ones in the amplitude analysis. The BF is given by

\begin{eqnarray} \begin{aligned}
    \mathcal{B}_{\text{sig}}=\frac{N_{\text{net}}}{\sum_{\alpha} N_{\alpha}^{\text{ST}}\epsilon^{\text{DT}}_{\alpha}/\epsilon_{\alpha}^{\text{ST}}},\, \label{eq:Bsig-gen}
\end{aligned} \end{eqnarray} 
where $\alpha$ runs over the various tag modes. The DT efficiency ($\epsilon_{\alpha}^{\rm DT}$) is the signal selection efficiency for an event with a $D^-$ in the $\alpha$-th tag mode. The ST yields ($N_{\alpha}^{\rm ST}$) and ST efficiencies ($\epsilon_{\alpha}^{\rm ST}$) for each tag mode are obtained by fitting the corresponding $M_{\rm BC}^{\rm tag}$ distributions individually. The corresponding $\epsilon_{\alpha}^{\rm DT}$, $N_{\alpha}^{\rm ST}$ and $\epsilon_{\alpha}^{\rm ST}$ are summarized in Table~\ref{tab:STeff}.   The DT yield $N_{\text{net}}$ is listed in Table~\ref{tab:sidebandBF}. 
In the fits to the $M_{\rm BC}^{\rm tag}$ distributions, the signal shape is modeled with the MC-simulated shape convolved with a double-Gaussian function. The background is
parameterized as an ARGUS function~\cite{argus} with an endpoint parameter fixed at 1.8865 GeV/$c^{2}$.   Figure~\ref{fig:DT_fit} shows the fit results. The candidates with $M_{\rm BC}^{\rm tag}$ within (1.863, 1.877) GeV/$c^2$ for each tag mode are kept for further analysis.
The corresponding DT efficiencies $\epsilon_{\alpha,\rm sig}^{\rm DT}$ are obtained by analyzing the signal MC samples, with the signal events for $D^+\to K_S^0K_S^0\pi^+$ generated based on the results of the amplitude analysis. 
After correcting for the differences in $\pi^{\pm}$ tracking, PID and $K_{S}^{0}$ reconstruction efficiencies between data and MC simulation, we determine the BF to be $(2.97 \pm 0.09_{\rm stat.} \pm 0.05_{\rm syst.}) \times 10^{-3}$. The systematic uncertainty will be discussed in the following section.
\begin{table}[hbtp]
  \begin{center}
    \caption{ The ST yields $(N_{\alpha }^{\rm ST})$, ST efficiencies $(\epsilon^{\rm ST}_{\alpha})$ and DT efficiencies $(\epsilon^{\rm DT}_{\alpha})$ for six tag modes. The BFs of the sub-particle ($K_{S}^{0}, \pi^{0}$) decays are not included. The uncertainties are statistical only.}
      \label{tab:STeff}
  \end{center}
    \begin{tabular}{lcccc }
      \hline
      \hline
      Tag mode $\alpha$   & $N_{\alpha}^{\rm ST}$ & {$\epsilon_{\alpha}^{\rm ST}(\%)$} &{$\epsilon_{\alpha}^{\rm DT}(\%)$} \\
      \hline
      $ K^+\pi^-\pi^-$         &2164074 $\pm$ 1571 & 51.17 $\pm$ 0.01  &11.00 $\pm$ 0.01\\  
      $ K^+\pi^-\pi^-\pi^0$    &\hspace{0.5em}689042  $\pm$ 1172 & 25.50 $\pm$ 0.01 & \hspace{0.5em}4.83 $\pm$ 0.01\\
      $ K_S^0\pi^-$            &250437  $\pm$ 524  & 50.63 $\pm$ 0.02 &10.83 $\pm$ 0.02\\
      $ K_S^0\pi^-\pi^0$       &558495  $\pm$ 930  & 26.28 $\pm$ 0.01 &\hspace{0.3em} 4.98 $\pm$ 0.01\\
      $ K_S^0\pi^-\pi^-\pi^+$  &300519  $\pm$ 669  & 28.97 $\pm$ 0.01 & \hspace{0.5em}4.94 $\pm$ 0.01\\
      $ K^+K^-\pi^-$           &187379  $\pm$ 541  & 41.06 $\pm$ 0.02 & \hspace{0.5em}8.75 $\pm$ 0.02\\
      
      \hline
      \hline
    \end{tabular}
    \end{table}

\section{VIII.~Systematic uncertainty OF BF MEASUREMENT}
Most systematic uncertainties related to the efficiency of reconstructing the $D^{\pm}$ mesons on the tag side are canceled due to the DT method. The following sources are taken into consideration to evaluate the systematic uncertainties in the BF measurement. The total systematic uncertainties are summarized in Table~\ref{BF-Sys}. Adding them in quadrature results in a total systematic uncertainty of 1.6\% in the BF measurement.

{(I) \textit{ST $D^-$ yield}:} the systematic uncertainty in the ST  yield has been estimated by varying the signal and background shapes in the fit. This systematic uncertainty is assigned to be 0.1\%.

{(II) \textit{Tracking and PID efficiencies}:} the data-MC efficiency ratios for $\pi^-$ tracking and PID efficiencies are determined to be $0.998 \pm 0.001$ and $0.998 \pm 0.001$ for this decay channel by using the same control samples used for the amplitude analysis uncertainties. After correcting the MC efficiencies to data by these factors, the statistical uncertainties of the correction parameters are assigned as the systematic uncertainties, which are 0.1\% for both $\pi^{\pm}$ PID and tracking.

{(III) \textit{2D fit}:} the uncertainty associated with the signal and background shapes in the 2D fit to the
DT $M_{\rm BC}$ distribution is 1.1\%, determined by varying the mean value and the resolution of the smeared Gaussian function by $\pm 1\sigma$ and varying the ARGUS end-point by $\pm 0.2$ MeV$/c^2$.

{(IV) \textit{$K_S^0$ reconstruction}:} the data-MC efficiency ratio for $K_S^0$ reconstruction is determined to be $0.990 \pm 0.004$ by using the same control samples used for the amplitude analysis uncertainties. After correcting the efficiency by this factor for the two $K_{S}^{0}$ mesons, we assign  0.8\% as the systematic uncertainty from the reconstruction of each $K_{S}^{0}$.

{(V) \textit{Amplitude model}:} the uncertainty from the signal MC model based on
the results of the amplitude analysis is studied by varying the fit
parameters according to the covariance matrix. The change of signal
efficiency is estimated to be 0.2\%.

{(VI) \textit{Peaking background estimation}:} according to the Eq.~(\ref{eq:net}), in the nominal analysis, the normalization factor for the peaking backgrounds (i.e. the ratio of the  background yields between the $K_{S}^{0}$ signal and the sideband 1 regions) is assumed to be 0.5. The BF is recalculated with alternative normalization factors determined by MC simulation. The corresponding change on the BF, 0.7\% for $D^{+} \to K_{S}^{0}K_{S}^{0}\pi^{+}$, is assigned as the systematic uncertainty associated with the peaking background normalization.  We have also examined the contributions of the sideband 2 region, but since the effects on the net number of events are negligible, we ignore them in this work.

{(VII) \textit{Quoted $\mathcal{B}$}:} in this measurement, the BF of $K_S^0 \to \pi^+\pi^-$ is quoted from the PDG~\cite{PDG}, which is $\mathcal{B}(K^{0}_{S}\to \pi^{+}\pi^{-})=(69.20\pm 0.05)\%.$ The associated uncertainty is assigned to be 0.1\%.  

{(VIII) \textit{$\Delta E$ requirement}:} considering the possible difference between data and MC simulation, we examine the $\Delta E_{\rm sig}$ cut efficiency after smearing a double-Gaussian function for signal MC sample, and we take the change of this efficiency,  0.5\%, as the systematic uncertainty.

\begin{table}[htbp]
  \caption{Systematic uncertainties for the BF measurement.}
  \centering
  \begin{tabular}{cc}
    \hline
    \hline
    Source   & Uncertainty~(\%) \\
    \hline
    ST $D^-$ yield  		&0.1\\
    PID efficiency            	    &0.1\\
    Tracking efficiency	        	&0.1\\
    2D fit        		            &1.1\\
    $K_S^0$ reconstruction 	        &0.8\\
    Amplitude  model            	&0.2\\
    Peaking background estimation   &0.7\\
    Quoted $\mathcal{B}$	        &0.1\\
    $\Delta E$ requirement          &0.5\\
    \hline
    Total    	 	                & 1.6\\
    \hline
    \hline
  \end{tabular}
  \label{BF-Sys}
\end{table}

\section{IX.~Summary}
Based on $e^{+}e^{-}$ collison data corresponding to an integrated luminosity of 7.93 fb$^{-1}$ collected with the BESIII detector at the center-of-mass energy 3.773~GeV, the first amplitude analysis of the decay $D^{+} \to K_{S}^{0}K_{S}^{0}\pi^{+}$ is performed. The fit results are listed in Table~\ref{fit-result}. In addition, with the detection efficiency obtained from the  signal MC sample generated based on the amplitude analysis results, we obtain the BF $\mathcal{B}(D^{+} \to K^{0}_{S}K^{0}_{S}\pi^{+})=(2.97\pm 0.09_{\rm stat.} \pm 0.05_{\rm syst.}) \times 10^{-3}$. The BF for $D^{+} \to K^{0}_{S}K^{*}(892)^{+}$ is calculated as
${\rm FF}\times \mathcal{B}(D^{+} \to K_{S}^{0}K_{S}^{0}\pi^{+})/\mathcal{B}(K^{*}(892)^{+} \to K_{S}^{0}\pi^{+})$ to be
$\mathcal{B}(D^{+} \to K^{0}_{S}K^{*}(892)^{+}) = (8.72 \pm 0.28_{\rm stat.} \pm 0.15_{\rm syst.}) \times 10^{-3}$,
which is consistent with the previous BESIII measurement~\cite{xxh} but with improved precision.

Compared to the theoretical expectations of Table~\ref{Consistentency_Check},  this result  deviates from those in Refs.~\cite{2,4} by more than five standard deviations, while is consistent with that based on the theoretical expectations~\cite{1,06316} within two standard deviations and  the  predictions in Ref.~\cite{3} within  three standard deviations.  

\section{ACKNOWLEDGEMENT}
\acknowledgements
The BESIII Collaboration thanks the staff of BEPCII and the IHEP computing center for their strong support. This work is supported in part by National Key R\&D Program of China under Contracts Nos. 2023YFA1606000,  2020YFA0406400, 2020YFA0406300; National Natural Science Foundation of China (NSFC) under Contracts Nos. 11635010, 11735014, 11935015, 11935016, 11935018, 12025502, 12035009, 12035013, 12061131003, 12192260, 12192261, 12192262, 12192263, 12192264, 12192265, 12221005, 12225509, 12235017, 12361141819, 12375087, 18175054; the Chinese Academy of Sciences (CAS) Large-Scale Scientific Facility Program; the CAS Center for Excellence in Particle Physics (CCEPP); Joint Large-Scale Scientific Facility Funds of the NSFC and CAS under Contract No. U1832207, U2032104; the Excellent Youth Foundation of Henan Scientific Committee under Contract No.~242300421044; 100 Talents Program of CAS; The Institute of Nuclear and Particle Physics (INPAC) and Shanghai Key Laboratory for Particle Physics and Cosmology; German Research Foundation DFG under Contracts Nos. 455635585, FOR5327, GRK 2149; Istituto Nazionale di Fisica Nucleare, Italy; Ministry of Development of Turkey under Contract No. DPT2006K-120470; National Research Foundation of Korea under Contract No. NRF-2022R1A2C1092335; National Science and Technology fund of Mongolia; National Science Research and Innovation Fund (NSRF) via the Program Management Unit for Human Resources \& Institutional Development, Research and Innovation of Thailand under Contract No. B16F640076; Polish National Science Centre under Contract No. 2019/35/O/ST2/02907; The Swedish Research Council; U. S. Department of Energy under Contract No. DE-FG02-05ER41374.

\end{document}